\documentclass[aps,superscriptaddress]{revtex4-2}

\usepackage{physics}
\usepackage{amsmath,amsfonts,amssymb}
\usepackage{braket}
\usepackage{mhchem}
\usepackage{bm}
\usepackage{nicefrac}
\usepackage{tikz}
\usetikzlibrary{quantikz}
\usetikzlibrary{calc}
\usepackage{yquant}
\usepackage{mathrsfs}
\usepackage{graphicx}
\usepackage[acronym]{glossaries}
\glsdisablehyper
\usepackage{float}
\usepackage{cleveref}
\crefname{figure}{Fig.}{Figs.}
\usepackage{natmove}
\usepackage{comment}
\usepackage[title]{appendix}
\usepackage[ruled,lined]{algorithm2e}
\usepackage[caption=false]{subfig}

\newacronym{PES}{PES}{potential energy surface}
\newacronym{MCTDH}{MCTDH}{multiconfiguration time-dependent Hartree}
\newacronym{GWP}{GWP}{Gaussian wavepacket}
\newacronym{GF}{GF}{Gaussian function}
\newacronym{FG}{FG}{frozen Gaussian}
\newacronym{vMCG}{vMCG}{variational multi-configurational Gaussian}
\newacronym{TDSE}{TDSE}{time-dependent Schr{\"o}dinger equation}
\newacronym{NISQ}{NISQ}{noisy intermediate-scale quantum}
\newacronym{VQA}{VQA}{variational quantum algorithm}
\newacronym{DV}{DV}{discrete variable}
\newacronym{CV}{CV}{continuous variable}
\newacronym{HDC}{HDC}{hybrid discrete-continuous}
\newacronym{CD}{CD}{conditional displacement}
\newacronym{EOM}{EOM}{equations of motion}
\newacronym{VP}{VP}{variational principle}
\newacronym{CV-DV}{CV-DV}{continuous-variable discrete-variable}
\newacronym{ECD}{ECD}{echoed conditional displacement}
\newacronym{non-QND}{non-QND}{not quantum non-demolition}

\begin{document}

\title{A variational hybrid continuous-variable discrete-variable quantum algorithm for adiabatic nuclear dynamics}

\author{Rami Gherib}
\affiliation{OTI Lumionics Inc., 3415 American Drive Unit 1, Mississauga, Ontario L4V\,1T4, Canada}

\author{Roya Radgohar}
\affiliation{Nord Quantique, 1950 rue Roy, Sherbrooke, Quebec J1K\,1B7, Canada}

\author{Anastasiia Pusenkova}
\affiliation{Nord Quantique, 1950 rue Roy, Sherbrooke, Quebec J1K\,1B7, Canada}

\author{Seyyed Mehdi Hosseini Jenab}
\affiliation{OTI Lumionics Inc., 3415 American Drive Unit 1, Mississauga, Ontario L4V\,1T4, Canada}

\author{Sara Turcotte}
\affiliation{Nord Quantique, 1950 rue Roy, Sherbrooke, Quebec J1K\,1B7, Canada}

\author{Dany Lachance-Quirion}
\affiliation{Nord Quantique, 1950 rue Roy, Sherbrooke, Quebec J1K\,1B7, Canada}

\author{Nicholas E. Frattini}
\affiliation{Nord Quantique, 1950 rue Roy, Sherbrooke, Quebec J1K\,1B7, Canada}

\author{Scott N. Genin}
\affiliation{OTI Lumionics Inc., 3415 American Drive Unit 1, Mississauga, Ontario L4V\,1T4, Canada}

\begin{abstract}
	\Gls{GWP} methods are prevalent means of solving the nuclear \gls{TDSE} on classical computers. They consist of representing the nuclear wave function as a superposition of Gaussians. Here, we present a variational hybrid \gls{CV-DV} algorithm for simulating adiabatic nuclear dynamics on qubit-oscillator quantum hardware. It encodes a superposition of \glspl{FG} onto the qubit-oscillator register and evolves it according to the time-dependent \gls{VP}. We test two variants of the approach, one that evolves a single \gls{FG} and another that evolves a superposition of \glspl{FG}, on a set of prototypical 1D harmonic and anharmonic systems. We simulate the photoexcited state dynamics of \ce{SO2} using the single \gls{FG} variant of the algorithm and compute its autocorrelation function on physical quantum hardware. For harmonic and Morse potentials, the variant evolving the superposition of \glspl{FG} converges to the numerically exact solution as the number of Gaussians increases. In the case of the double-well potential, the algorithm simulates wavepacket bifurcation and recurrence correctly, and the ensuing irregular dynamics moderately well. Overall, this work establishes a pathway for simulating molecular dynamics on noisy intermediate-scale quantum devices.
\end{abstract}

\maketitle

\setlength{\parskip}{7pt}

\glsresetall

\section{Introduction}
Solving the \gls{TDSE} is necessary for understanding and simulating many chemical processes. Unfortunately, numerically exact calculations typically scale exponentially with the number of nuclear degrees of freedom and become prohibitively resource-consuming for large molecules~\cite{kosloff1988time, beck2000multiconfiguration}. In many cases, however, it suffices to use approximative methods for quantum dynamics. A notable example are \gls{GWP} methods, which represent the nuclear wave function as a finite superposition of Gaussian functions. 
Their objective is to evolve the wave function parameters, such as the expansion coefficient, position, momentum and phase of each individual Gaussian, so as to approximate exact quantum dynamics. Different \gls{GWP} methods differ in part in how their parameters evolve. 
Notably are the ones that evolve all their parameters variationally~\cite{sawada1985strategy, sawada1986gaussian, richings2015quantum, worth2003full, mendive2012towards, worth2020gaussian}. A fully variational approach assures that dynamics converge to the exact result as we increase the number of Gaussians and, conceptually, solves the \gls{TDSE} exactly in the limit of infinite Gaussians~\cite{mendive2012towards}. However, one limitation is that the number of Gaussians required to reach high levels of accuracy often increases rapidly with the number of spatial dimensions. Although the scaling is highly system-dependent, in principle it can approach exponential scaling if a full basis is required. 

The considerable computational costs needed to perform quantum dynamics simulations of large molecules have motivated researchers to pursue alternative computational paradigms, namely quantum computations. Quantum computers can potentially perform exact quantum simulations with computational costs that scale polynomially with respect to the size of the system.~\cite{feynman2018simulating, tacchino2020quantum, miessen2023quantum} 
\Glspl{VQA} have been proposed as a means for obtaining quantum advantage on \gls{NISQ} computers~\cite{cerezo2021variational}. These algorithms involve solving a problem by representing its approximate solution on a quantum circuit. By iteratively adjusting the circuit parameters to minimize a cost function, the approximate solution improves until it becomes nearly optimal. 

In recent years, \glspl{VQA} have been developed for nuclear dynamics simulations~\cite{lee2022variational, ollitrault2021molecular, ollitrault2023quantum}. These consist of representing the nuclear wave function as the state of a multi-qubit register. The wave function is parametrically dependent on the gate parameters whose circuit architecture is defined heuristically. These evolve in accordance to the McLachlan \gls{VP}~\cite{mclachlan1964variational} producing the optimal time-evolution of the nuclear wave function.
The possible quantum advantage stems from needing a smaller number of parameters to represent the wave function as the state of a quantum register compared to its representation on classical computers~\cite{ollitrault2021molecular, li2017efficient}.

Conventionally, quantum algorithms for nuclear dynamics have focused on the \gls{DV} model~\cite{wiesner1996simulations, zalka1998simulating, somma2003quantum, kassal2008polynomial, ollitrault2020nonadiabatic}, where quantum information is encoded in qubits. The use of the \gls{DV} model, however, comes with significant computational costs~\cite{liu2026hybrid}, which stems in part from nuclear wave functions belonging to infinite-dimensional Hilbert spaces, while multi-qubits states belong to finite-dimensional ones. It is in part for this reason that quantum algorithms based on the \gls{CV} model have been presented as an alternative~\cite{macdonell2021analog, liu2026hybrid, malpathak2025simulating}. The \gls{CV} quantum computing model encodes quantum information in continuous degrees of freedom. Its fundamental unit of information is the \emph{qumode} whose states exist in an infinite-dimensional Hilbert space. 
A qumodes is typically a single bosonic mode (harmonic oscillator), and its state is manipulated using Gaussian and non-Gaussian operations. Compared to qubits, they typically have longer lifetimes and are more naturally suited for representing nuclear wave functions, since the dimensionality of their Hilbert space matches that of nuclear wave functions~\cite{macdonell2021analog, malpathak2025simulating, liu2026hybrid, macdonell2023predicting, valahu2023direct, navickas2025experimental}. 

Hybrid qubit-oscillator architectures coherently couple qubits to qumodes~\cite{liu2026hybrid}. They have recently gained attention as viable platforms for developing and executing new quantum algorithms and simulations. Hybrid \gls{CV-DV} algorithms involve \gls{DV}, \gls{CV} and qubit-controlled operations. The latter are operations applied to qumodes that depends on the state of the qubit and generate entanglement between the qubit and the oscillator. Hybrid \gls{CV-DV} algorithms can be experimentally realized using 3D coaxial cavities coupled to transmons, trapped-ion, and neutral-atom platforms~\cite{liu2026hybrid}.

This work presents a proof-of-concept variational hybrid \gls{CV-DV} algorithm for simulating adiabatic nuclear dynamics.
The \gls{VQA} uses qubit rotation gates and \gls{CD} gates to embed a superposition of squeezed states/\glspl{FG} representing the nuclear wave function onto the register. The \gls{VQA} is effectively a \gls{GWP} method where the variational parameters are the qubit rotation angles, controlling the phases and amplitudes of the individual Gaussians, and the \gls{CD} parameters, controlling their positions and momenta, evolving according to the Kramer-Saraceno \gls{VP}~\cite{kramer1981geometry}. The number of Gaussians embedded in the register is controlled by the number of ancilla qubits. With each additional pair of ancilla qubits, the number of Gaussians in the superposition doubles. Consequently, the number of Gaussians scales exponentially with the allotted computational resources. Our work can be viewed as hybrid \gls{CV-DV} approach of the algorithm by Ollitrault~\textit{et~al.}~\cite{ollitrault2023quantum} based on the \gls{DV} paradigm. A key approximation of the algorithm is that the initial wavefunction is Gaussian; a typical scenario in many photochemical processes, as molecules are near their equilibrium where the \gls{PES} is nearly quadratic.

The remainder of this manuscript starts by presenting a simpler version of the algorithm in which the vibrational wavepacket is approximated to be a single \gls{FG}. Then, it presents the complete algorithm for dynamics of a superposition of Gaussians. Numerical calculations are presented to assess the two variants of the algorithm on simple molecular models, namely quadratic, Morse and double-well potentials. The photoexcited state dynamics of \ce{SO2} are simulated using the single \gls{FG} variant. Its variational parameters are then used to compute the autocorrelation function on a quantum processor comprised of a single-mode 3D coaxial cavity dispersively coupled to a transmon. Then we assess whether the \gls{VQA} propagating a superposition of frozen Gaussians can simulate thawed Gaussian dynamics on a quadratic model. This is a relatively easy benchmark test. The Morse potential represents a test of moderate difficulty that assesses the \gls{VQA}'s ability to model localized anharmonic dynamics. The double-well potential represents a greater challenge in assessing whether the method can simulate quantum effects such as wavepacket bifurcation, recurrence and coherent interference~\cite{begusic2022applicability, ryabinkin2024thawed, sharma2026variational}. In the case of harmonic and Morse potentials, the \gls{VQA} gradually converges to the numerically exact solutions when the number of ancilla qubits increases. In the case of a double-well potential, the algorithm can simulate wavepacket bifurcation and recurrence correctly, and the ensuing irregular dynamics moderately well. Finally, we conclude by discussing the limitations of the algorithm. We note that our investigation pertains solely to one-dimensional models where the core properties of the algorithm can be highlighted. These models serve as a natural first benchmark before extending to higher-dimensional systems.

%%%%%%%%%%%%%%%%%%%%%%%%%%%%%%%%%%%%%%%%%%%%%%%%%%%%%%%%%%
%%%%%%%%%%%%%%%%%%%%%%%%%%%%%%%%%%%%%%%%%%%%%%%%%%%%%%%%%%

\section{Quantum algorithm}

\subsection{\gls{VQA} variant of dynamics of a single \gls{FG}}

\label{sec:var_dyn_single}

Before presenting the \gls{VQA} for a superposition of \glspl{FG}, we introduce the simpler version of the algorithm for a single \gls{FG}. It is appropriate to begin with this relatively simple case because the algorithm of the superposition case is an extension of the single \gls{FG} one. We also preface that while the dynamics of a superposition of Gaussians satisfy the Kramer-Saraceno \gls{VP}, the single \gls{FG} case utilizes the McLachlan \gls{VP}. The reasons for choosing different \glspl{VP} are given in Appendix~\ref{subsec:norm-conserv}.

A single \gls{FG} in 1D $\Psi_{G}\left(r\right)$ is parametrized by its center in phase space $\left[x_{t}, p_{t}\right]$ and its global phase $\gamma_{t}$,
\begin{equation}
\label{eq:1d_gaussian}
\Psi_{G}\left(r|x_{t}, p_{t}, \gamma_{t}\right) = N_{G}\exp\left(-\sigma\left(r-x_{t}\right)^{2}+ip_{t}\left(r-x_{t}\right)+i\gamma_{t}\right)
,
\end{equation}
where $N_{G}$ is its normalization constant, $x_{t}$ is the average position and $p_{t}$ is the average momentum. The parameter $\gamma_{t}$ is complex and ensures norm conservation during variational dynamics~\cite{lasser2022various}. The parameters $x_{t}$, $p_{t}$ and $\gamma_{t}$ evolve in time and, in the context of simulations on classical computers, their \gls{EOM} are  well-known~\cite{coalson1990multidimensional}. 

A register composed of one qubit and one qumode is typically initialized to $\ket{\Psi_{0}}=\ket{0}\ket{0}_{m}$, where $\ket{0}$ is the qubit ground state and $\ket{0}_{m}$ is the vacuum state of the qumode. The information present in Eq.~\eqref{eq:1d_gaussian} can be encoded onto the state of the register. Gaussian gates, in particular the pure position displacement gate, $\hat{X}\left(x\right)=\exp\left(-ix\hat{p}/\hbar\right)$, and the pure momentum displacement gate $\hat{Z}\left(p\right)=\exp\left(ip\hat{x}/\hbar\right)$, control the position of the state of the qumode in phase space and its phase through its geometric phase, while a qubit rotation gate $\hat{R}_{z}\left(\gamma\right)$ controls the phase directly. The state of the oscillator-qubit register $\ket{\Psi(\bm{\lambda_{G}})} = \hat{R}_{z}\left(\gamma\right)\ket{0}\hat{Z}\left(p\right)\hat{X}\left(x\right)\ket{0}_{m}$ where $\bm{{\lambda}_{G}} = \left[{x}, {p}, {\gamma}\right]$ embeds the same information present in $\Psi_{G}$ in Eq.~\ref{eq:1d_gaussian}.  We denote the unitary that acts solely on the qumode as $\hat{G}\left(x, p \right) = \hat{Z}\left(p\right)\hat{X}\left(x\right)$ and the unitary that acts on both the qubit and the qumode as $\hat{U}\left(\bm{\lambda_{G}} \right) = \hat{R}_{z}\left(\gamma\right)\hat{G}\left(x, p \right)$. 

The McLachlan \gls{VP}~\cite{mclachlan1964variational} approximates the time evolution of the variational parameters $\bm{\lambda}_{G}$
under the nuclear Hamiltonian $\hat{\mathcal{H}}$, giving the \gls{EOM} 
\begin{equation}
\label{eq:MVP}
\text{Re}\left(\bm{M}\right)\bm{\dot{\lambda}_{G}}=\frac{1}{\hbar}\text{Im}(\bm{V}),
\end{equation}
where $\bm{\dot{\lambda}_{G}} = \left[\dot{x}, \dot{p}, \dot{\gamma}\right]$ is an array that contains the time-derivatives of the variational parameters, ${M}_{jk}=\left\langle\frac{\partial\Psi(\bm{\lambda_{G}})}{\partial \lambda_{{G}_{j}}}\middle|\frac{\partial\Psi(\bm{\lambda_{G}})}{\partial \lambda_{{G}_{k}}}\right\rangle$ and $V_{j} = \left\langle\frac{\partial\Psi(\bm{\lambda_{G}})}{\partial \lambda_{{G}_{j}}}\middle|\hat{\mathcal{H}}\middle|\Psi(\bm{\lambda_{G}})\right\rangle$.
 The variational dynamics of a single \gls{FG} are summarized in Algorithm~\ref{alg:1FG_var_qdyn} wherein $\text{Re}\left(\bm{M}\right)$ and  $\text{Im}\left(\bm{V}\right)$ are evaluated at each time step, and $\bm{\lambda_{G}}$ is propagated numerically. 
\\
\begin{algorithm}[H]
\label{alg:1FG_var_qdyn}
\caption{Variational dynamics of a single \gls{FG}}
\KwIn{Initial set of parameters 
$\bm{\lambda_{G}}(0) = \left[x(0), p(0), \gamma(0) \right]$, 
}
\KwOut{Trajectories in parameter space $\bm{\lambda_{G}}(t) = \left[x(t), p(t), \gamma(t)\right]$}
Set
$\bm{\lambda_{G}} \gets \bm{\lambda_{G}}(0)$ \;
\For{$t = 0$ \KwTo $T-1$}
{
    Compute $\text{Re}(\bm{M}) \gets \text{Re}(\bm{M}(\bm{\lambda_{G}}))$ \;
    Compute $\text{Im}(\bm{V}) \gets \text{Im}(\bm{V}(\bm{\lambda_{G}}))$ \;
    Solve for $\dot{\bm{\lambda}}_{\bm{G}}$ in $\text{Re}(\bm{M})\dot{\bm{\lambda}}_{\bm{G}}=\frac{1}{\hbar}\text{Im}(\bm{V})$\;
    Update $\bm{\lambda_{G}} \gets \textsc{Integrate}(\bm{\lambda_{G}}, \dot{\bm{\lambda}}_{\bm{G}}, \Delta t)$\;
    Store $\bm{\lambda_{G}}$\;
}
\Return $\bm{\lambda_{G}}(t)$\;
\end{algorithm}

The task at hand is to construct \gls{CV-DV} quantum circuits where expectation values correspond to the matrix elements of $\text{Re}\left(\bm{M}\right)$ and $\text{Im}\left(\bm{V}\right)$. We approximate the derivatives using finite central differences 
\begin{equation}
\left|{\frac{\partial \Psi\left(\bm{\lambda_{G}}\right)}{\partial \lambda_{G_{j}}}}\right\rangle \approx \frac{\hat{U}\left(\lambda_{G_{j}}+\epsilon\right)\ket{\Psi_0} - \hat{U}\left(\lambda_{G_{j}} - \epsilon\right)\ket{\Psi_0}}{2\epsilon},
\label{eq:FCD}
\end{equation}
where $\epsilon$ is an arbitrarily small number. This requires embedding multiple derivatives of $\ket{\Psi\left(\bm{\lambda_{G}}\right)}$ simultaneously within the \gls{CV-DV} circuit. 

Although an exact scheme, such as one involving the parameter-shift rule, is plausible, we have not attempted it; our investigation is exploratory and parameter-shift rules for \gls{CV-DV} generators are underdeveloped. While their derivations are certainly worthwhile, they lie outside the scope of our current work and may require a substantial digression. We have opted to verify the functionality of our overall approach and leave the optimization of its different parts to future investigations. 

Using Eq.~\ref{eq:FCD}, we can approximate $\text{Re}\left(M_{jk}\right)$ and $\text{Im}\left(V_{j}\right)$ as
\begin{equation}
\label{eq:Mjk_approx_gauss}
\text{Re}\left(M_{jk}\left(\bm{\lambda_{G}}\right)\right)
\approx 
\frac{\text{Re}\left(
\left\langle{\hat{U}\left(\lambda_{G_{j}}+\epsilon_{j}\right)\Psi_{0} - \hat{U}\left(\lambda_{G_{j}}-\epsilon_{j}\right)\Psi_{0} 
\Big| 
\hat{U}\left(\lambda_{G_{k}}+\epsilon_{k}\right)\Psi_{0} - \hat{U}\left(\lambda_{G_{k}}-\epsilon_{k}\right)\Psi_{0}}
\right\rangle\right)}{4\epsilon_{j}\epsilon_{k}}
\end{equation}
and
\begin{equation}
\label{eq:Vj_approx_gauss}
\text{Im}\left(V_{j}\left(\bm{\lambda_{G}}\right)\right) 
\approx 
\frac{\text{Im}\left(
\left\langle
{\hat{U}\left(\lambda_{G_{j}}+\epsilon_{j}\right)\Psi_{0} - \hat{U}\left(\lambda_{G_{j}}-\epsilon_{j}\right)\Psi_{0}
\Big|
\hat{\mathcal{H}}
\Big|
\hat{U}\left(\bm{\lambda_{G}}\right)\Psi_{0}}
\right\rangle
\right)}{2\epsilon_{j}}.
\end{equation}

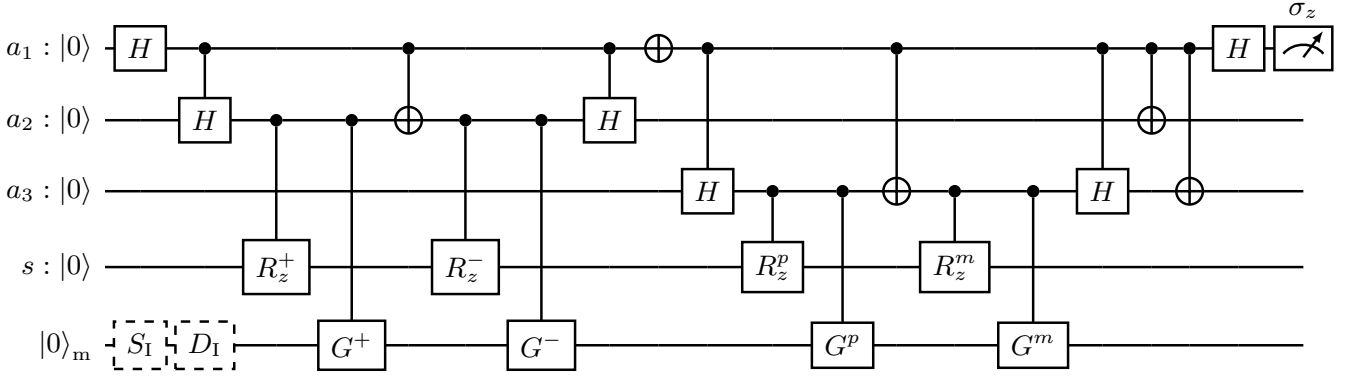
\begin{figure}[H]\centering
	\resizebox{1.0\textwidth}{!}{
\begin{quantikz}[row sep=0.3 cm, column sep=0.1cm]
\lstick{$a_{1}: \ket{0}$} &\gate{H} &\ctrl{1}& \qw              & \qw          &\ctrl{1} &\qw              &\qw          &\ctrl{1} &\targ{} &\ctrl{2} &\qw              &\qw 
&\ctrl{2} &\qw          &\qw          &\ctrl{2} &\ctrl{1} &\ctrl{2} &\gate{H} &\meter{$\sigma_{z}$}\\
\lstick{$a_{2}: \ket{0}$} &\qw      &\gate{H}& \ctrl{2}         & \ctrl{3}     &\targ{}  &\ctrl{2}         &\ctrl{3}     &\gate{H} &\qw     &\qw      &\qw              &\qw 
&\qw      &\qw          &\qw          &\qw      &\targ{}  &\qw     &\qw &\qw\\
\lstick{$a_{3}: \ket{0}$} &\qw      &\qw     & \qw              & \qw          &\qw      &\qw              &\qw          &\qw      &\qw     &\gate{H} &\ctrl{1}         &\ctrl{2} 
&\targ{}  &\ctrl{1}     &\ctrl{2}     &\gate{H} &\qw      &\targ{} &\qw &\qw\\
\lstick{$s: \ket{0}$}     &\qw      &\qw     & \gate{R_{z}^{+}} & \qw          &\qw      &\gate{R_{z}^{-}} &\qw          &\qw      &\qw     &\qw      &\gate{R_{z}^{p}} &\qw 
&\qw      &\gate{R_{z}^{m}} &\qw          &\qw     &\qw       &\qw     &\qw &\qw\\
\lstick{$\ket{0}_{\text{m}}$} & \gate[style={dashed}]{S_{\text{I}}}     & \gate[style={dashed}]{D_{\text{I}}}     & \qw              & \gate{G^{+}} &\qw      &\qw              &\gate{G^{-}} &\qw      &\qw     &\qw      &\qw              &\gate{G^{p}} 
&\qw      &\qw          &\gate{G^{m}} &\qw     &\qw       &\qw     &\qw &\qw
\end{quantikz}
	}
\caption{The \gls{CV-DV} circuit for approximating $\text{Re}\left(M_{jk}\right)$ for single \gls{FG} dynamics where $U(\lambda_{G_{j}}+\epsilon_{j}) = R_{z}^{+}G^{+}$, $U(\lambda_{G_{j}}-\epsilon_{j}) = R_{z}^{-}G^{-}$, $U(\lambda_{G_{k}}+\epsilon_{k}) = R_{z}^{p}G^{p}$ and $U(\lambda_{G_{k}}-\epsilon_{k}) = R_{z}^{m}G^{m}$. Since $\text{Re}(\bm{M})$ is symmetric, the circuit only needs to run with unordered pairs of $j$ and $k$. Ancilla qubits are denoted $a_{n}$ and the so-called system qubit that encodes the phase information is denoted $s$. The numerator in Eq.~\eqref{eq:Mjk_approx_gauss} is approximated by $\braket{\sigma_{z}}$. The dashed $S_{\text{I}}$ and $D_{\text{I}}$ gates are optional depending on how $\mathcal{H}$ measurements are performed in Fig.~\ref{fig:Vj_gauss_circuit}, see Appendix~\ref{sec:H_measmt}. The $H$ gate is the Hadamard gate, $H =\frac{1}{\sqrt{2}} \begin{bsmallmatrix} 1 & 1 \\ 1 & -1 \end{bsmallmatrix}$.}
\label{fig:Mjk_gauss_circuit}
\end{figure}

\begin{figure}[H]\centering
	\resizebox{0.8\textwidth}{!}{
\begin{quantikz}[row sep=0.3 cm, column sep=0.1cm]
\lstick{$a_{1}: \ket{0}$}       & \gate{H} & \ctrl{1} & \qw                  & \qw      & \ctrl{1} 
& \qw                  & \qw           & \ctrl{1} & \targ{} & \ctrl{2}             
& \ctrl{3}    & \ctrl{1} & \gate{H} & \meter{$\sigma_{z}$} \\
\lstick{$a_{2}: \ket{0}$}       & \qw      & \gate{H} & \ctrl{1}             & \ctrl{2} & \targ{} 
& \ctrl{1}             & \ctrl{2}      & \gate{H} & \gate{S}   & \qw                  
& \qw         & \targ{}    & \qw      & \qw  \\ 
\lstick{$s: \ket{0}$}       & \qw      & \qw      & \gate{R_{z}^{+}} & \qw      & \qw 
& \gate{R_{z}^{-}} & \qw           & \qw      & \qw   & \gate{R_{z}^{0}} 
& \qw         & \qw      & \qw      & \qw  \\
\lstick{$\ket{0}_{\text{m}}$}& \gate[style={dashed}]{S_{\text{I}}}      & \gate[style={dashed}]{D_{\text{I}}}      & \qw                  & \gate{G^{+}} & \qw
& \qw                  & \gate{G^{-}} & \qw      & \qw   & \qw                   & \gate{G^{0}} & \qw      & \qw      & \meter{$\mathcal{H}$} 
\end{quantikz}
	}
\caption{The \gls{CV-DV} circuit for approximating $\text{Im}\left(V_{j}\right)$ for single \gls{FG} dynamics where $U(\lambda_{G_{j}}+\epsilon_{j}) = R_{z}^{+}G^{+}$, $U(\lambda_{G_{j}}-\epsilon_{j}) = R_{z}^{-}G^{-}$ and $U(\bm{\lambda_{G}}) = R_{z}G^{0}$. The numerator in Eq.~\eqref{eq:Vj_approx_gauss} is approximated as $\mathbb{E}\left[\left(-1\right)^{f}\mathcal{H}_{m}\right]$ where $f=0$ if $a_{1}$  collapses to $\ket{0}$ or $f=1$ if it collapses to $\ket{1}$, and $\mathcal{H}_{\text{m}}$ is the m\textsuperscript{th} measurement of $\hat{\mathcal{H}}$~\cite{ollitrault2021molecular}. The dashed $S_{\text{I}}$ and $D_{\text{I}}$ gates are optional depending on how $\hat{\mathcal{{H}}}$ measurements are performed, see Appendix~\ref{sec:H_measmt}. The $S$ gate is the phase gate, $S =\begin{bsmallmatrix} 1 & 0 \\ 0 & i \end{bsmallmatrix}.$}
\label{fig:Vj_gauss_circuit}
\end{figure}
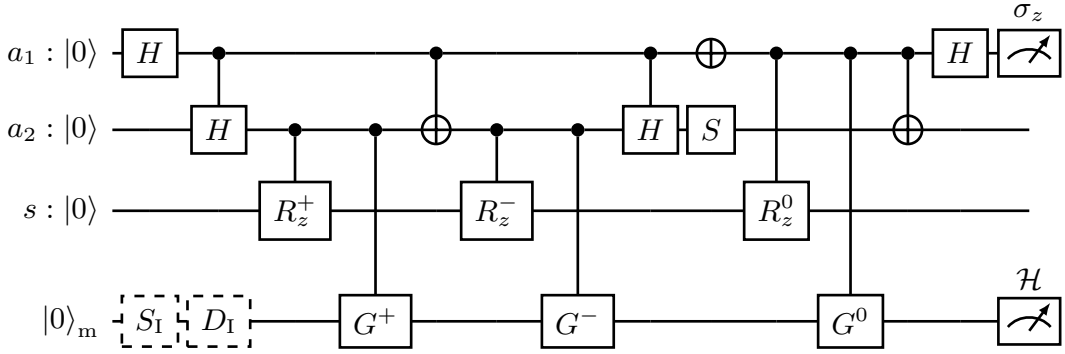

The quantum circuits used to compute the numerators in Eqs.~\eqref{eq:Mjk_approx_gauss} and \eqref{eq:Vj_approx_gauss} are constructed as follows. First, we begin with the qubit and qumode that embed a single \gls{FG} and introduce three additional ancilla qubits for the circuit approximating $\text{Re}(\bm{M})$ and two additional ancilla qubits for the circuit approximating $\text{Im}(\bm{V})$. \gls{CD} gates translate the qumode state in different directions in phase space, such that different ancilla qubits states end up being associated with \glspl{FG} with different small displacements in phase space. Hadamard gates are subsequently used to create a superposition of all displaced Gaussians. The circuit measurements are used to construct the numerators of Eqs.~\eqref{eq:Mjk_approx_gauss} and~\eqref{eq:Vj_approx_gauss}. Elements $\text{Re}\left(M_{jk}\right)$ are obtained by sampling the circuit in Fig.~\ref{fig:Mjk_gauss_circuit} and elements $\text{Im}\left(V_{j}\right)$ are evaluated by sampling the circuit in Fig.~\ref{fig:Vj_gauss_circuit}, measuring the ancilla qubit and the qumode. These circuits can be regarded as \gls{CV-DV} versions of the ones presented in Refs.~\citenum{lee2022variational, ollitrault2021molecular, ollitrault2023quantum} based on the \gls{DV} paradigm.

\subsection{\gls{VQA} variant of dynamics of a superposition of \glspl{FG}}
\label{sec:var_dyn_superpos}

This section introduces the \gls{VQA} for dynamics of a superposition of \glspl{FG}. A superpositions of \glspl{FG}, $\Psi_{\text{S}}(r)$, is often represented as~\cite{richings2015quantum}
\begin{equation}
\label{eq:mgwf_classical}
\Psi_{\text{S}}\left(r|\bm{\xi}(t), \bm{\eta}(t)\right) = \sum_{k=1}^{n} c_{k} \exp\left(\zeta_{k} r^2 + \xi_{k}(t) r + \eta_{k}(t) \right),
\end{equation}
where $\xi_{k}(t)$ and $\eta_{k}(t)$ are the time-dependent variational parameters and $\zeta_{k}$ is the width of the $k$\textsuperscript{th} Gaussian function. Collectively they determine the phase, the position and the momentum of each individual Gaussian.

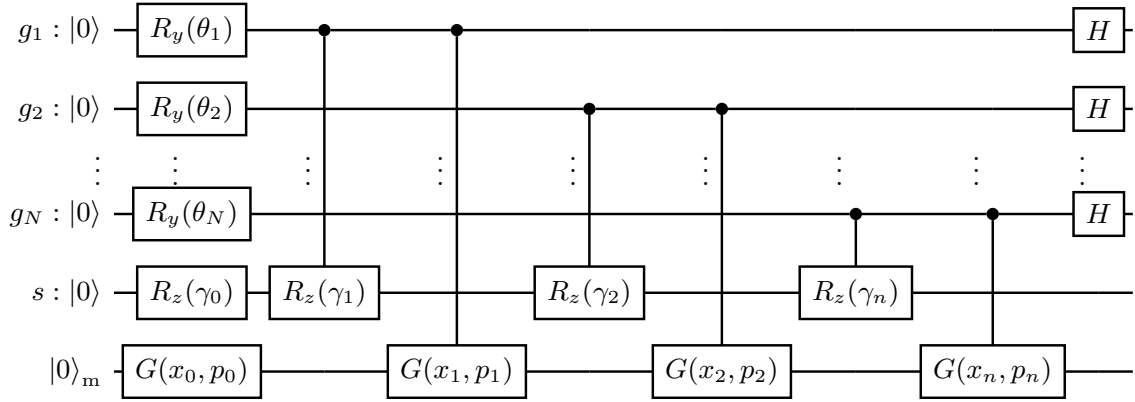
\begin{figure}[H]\centering
		\resizebox{0.85\textwidth}{!}{
\begin{quantikz}[row sep=0.3 cm, column sep=0.1cm]
\lstick{$g_{1}: \ket{0}$} & \gate{R_{y}(\theta_{1})} & \ctrl{4} & \ctrl{5}  
& \qw & \qw & \qw & \qw & \gate{H} & \qw \\
\lstick{$g_{2}: \ket{0}$} & \gate{R_{y}(\theta_{2})} & \qw & \qw   
& \ctrl{3} & \ctrl{4} & \qw & \qw & \gate{H} & \qw \\
\lstick{$\vdots$} & \lstick{$\vdots$} & \lstick{$\vdots$} & \lstick{$\vdots$}& \lstick{$\vdots$}& \lstick{$\vdots$}& \lstick{$\vdots$}& \lstick{$\vdots$}& \lstick{$\vdots$}& \\
\lstick{$g_{N}: \ket{0}$} & \gate{R_{y}(\theta_{N})} & \qw & \qw  
& \qw & \qw & \ctrl{1} & \ctrl{2} & \gate{H} & \qw \\
\lstick{$s: \ket{0}$} & \gate{R_{z}(\gamma_{0})} & \gate{R_{z}(\gamma_{1})} & \qw  
& \gate{R_{z}(\gamma_{2})} & \qw & \gate{R_{z}(\gamma_{n})} & \qw & \qw & \qw \\
\lstick{$\ket{0}_{\text{m}}$} & \gate{G(x_{0},p_{0})} & \qw & \gate{G(x_{1},p_{1})}  
& \qw & \gate{G(x_{2},p_{2})} & \qw & \gate{G(x_{n},p_{n})} & \qw &\qw 
\end{quantikz}
	}
\caption{\gls{CV-DV} circuit for embedding a superposition of $2^{N}$ Gaussians. Ancilla qubits are denoted as $g_{n}$ and the system qubit containing phase information is denoted as $s$.}
\label{fig:mgwp_circuit}
\end{figure}

A superposition of $2^{N}$ \glspl{FG} can be embedded onto an oscillator-qubits register comprised of 
%so that it contains the same information-type as $\Psi_{\text{S}}\left(r|\bm{\xi}, \bm{\eta}\right)$. 
a qumode, a system qubit, and $N$ ancilla qubits. This is done by assigning different coherent states to the different computational bases of the ancilla subsystem. The circuit that does so is depicted in Fig.~\ref{fig:mgwp_circuit}.  
It is initialized $\ket{0}^{\otimes N}\ket{0}\ket{0_{\text{m}}}$ before $N$ qubit rotation gates on the ancilla subsystem induce a superposition of ancilla computational basis states, all multiplied by $\ket{0}\ket{0_{\text{m}}}$. Gates $R_{z}(\gamma_{0})$ and $G(x_{0},p_{0})$ are then applied on the system qubit and qumode thereby assigning the same coherent state to all ancilla computational bases. Subsequently, the phases of the system qubit, and the positions and momenta of the coherent states associated to different computational bases are altered using controlled-$R_{z}$ and \gls{CD} gates; this effectively associates each computational basis with a different Gaussian function. 

Subsequently, we apply Hadamard gates on the ancilla qubits in the last column of Fig.~\ref{fig:mgwp_circuit}; the superposition of interest $\ket{\Psi(\bm{\lambda})}$ is then the projection of the full register state onto the $\ket{0}^{\otimes N}$ state of the ancilla subsystem. Effectively, the wave function ansatz $\ket{\Psi(\bm{\lambda})}$ of our algorithm is 
\begin{align}
\label{eq:mgwf_qregist_ansatz}
\ket{\Psi(\bm{\lambda})} &=
\frac{{1}}{2^{N/2}}
\sum_{k=0}^{2^{N}-1}
\left[
\prod_{n=1}^{N}
\left(\cos\left(\frac{\theta_{n}}{2}\right)\right)^{1-\text{bin}(k)_{n}}
\left(\sin\left(\frac{\theta_{n}}{2}\right)\right)^{\text{bin}(k)_{n}}
\right] \\
&\quad 
\left[\prod_{n=1}^{N}
\left(R_{z}\left(\gamma_{n}\right)\right)^{\text{bin}(k)_{n}}
\right]
R_{z}\left(\gamma_{0}\right)\ket{0} 
\left[
\prod_{n=1}^{N}
G\left(x_{n}, p_{n}\right)^{\text{bin}(k)_{n}}
\right]
G\left(x_{0}, p_{0}\right)\ket{0}_{\text{m}}, \nonumber 
\end{align}
where $\text{bin}(k)_{n}$ is the $n$\textsuperscript{th} digit of the binary representation of the number $k$. $\ket{\Psi(\bm{\lambda})}$ is a superposition of $2^{N}$ \glspl{FG} parameterized by the circuit gate parameters $\bm{\lambda} = [\bm{\theta}, \bm{x}, \bm{p}, \bm{\gamma}]$, where $\bm{\theta}=[\theta_{1}, ..., \theta_{N}]$, $\bm{x}=[x_{0}, ..., x_{N}]$, $\bm{p}=[p_{0}, ..., p_{N}]$ and $\bm{\gamma}=[\gamma_{0}, ..., \gamma_{N}]$. The rotation angles $\bm{\theta}$ control the weight of the individual \glspl{FG}, while the $R_{z}(\gamma_{n})$ and $G\left(x_{n}, p_{n}\right)$ gates control their phases and the centers in phase space. The $2^{-N/2}$ factor comes from the application of the $N$ Hadamard gates.

Intuitively, we can think of this wave function ansatz as follows. It is essentially the weighted sum of a \emph{guiding} Gaussian and $2^{N}-1$ \emph{ancillary} Gaussians. The phase and coordinates of the guiding Gaussian are $\gamma_0$, $x_{0}$ and $p_{0}$. When $\bm{\theta}=\bm{0}$, the weight of the guiding Gaussian is $1$ and the ancillary Gaussians do not contribute to the superposition. As $\bm{\theta}$ changes, population is transferred from the guiding Gaussian to the ancillary Gaussians. The parameters $\gamma_{n>0}$, $x_{n>0}$ and $p_{n>0}$ control the phases and centers of the ancillary Gaussians \emph{relative} to the guiding Gaussian. When all $\gamma_{n>0}$, $x_{n>0}$ and $p_{n>0}$ are $0$, the ancillary Gaussians have effectively the same phases and centers as the guiding Gaussian. Changing $x_{n>0}$ and $p_{n>0}$ translates them away from the center of the guiding Gaussian, while changing $\gamma_{n>0}$ induces a relative phase. $\ket{\Psi(\bm{\lambda})}$ contains the same information as $\Psi_{S}\left(x|\bm{\xi}, \bm{\eta}\right)$. Doubling the number of Gaussian functions in $\ket{\Psi(\bm{\lambda})}$ requires one additional set of ancilla qubit, controlled-$R_{z}$ gate, Hadamard gate and a pair of \gls{CD} gates. %Doubling the number of Gaussians in $\Psi_{S}\left(x|\bm{\xi}, \bm{\eta}\right)$ requires doubling the number of variational classical parameters.
 
The \gls{EOM} of $\bm{\lambda}$ are derived from the Kramer-Saraceno \gls{VP}~\cite{kramer1981geometry}
\begin{equation}
\label{eq:KSVP}
\text{Im}\left(\bm{M}\right)\bm{\dot{\lambda}}=-\frac{1}{\hbar}\text{Re}(\bm{V}),
\end{equation}
and contain the corresponding $\bm{M}$ and $\bm{V}$ arrays containing the partial derivatives of $\Psi(\bm{\lambda})$ with respect to the elements of $\bm{\lambda}$. Analogously to the single Gaussian case, the task at hand is to construct \gls{CV-DV} circuits where the expectation values are $\text{Im}(M_{jk}(\bm{\lambda}))=\text{Im}\left(\braket{\frac{\partial\Psi(\bm{\lambda})}{\partial \lambda_{j}}|\frac{\partial\Psi(\bm{\lambda})}{\partial \lambda_{k}}}\right)$ and $\text{Re}(V_{j}(\bm{\lambda})) = \text{Re}\left(\braket{\frac{\partial\Psi(\bm{\lambda})}{\partial \lambda_{j}}|\hat{\mathcal{H}}|\Psi(\bm{\lambda})}\right)$. 
The variational quantum dynamics simulation procedure is summarized in Algorithm~\ref{alg:var_qdyn} where $\text{Im}\left(\bm{M}\right)$ and $\text{Re}\left(\bm{V}\right)$ are evaluated at each time step, and $\bm{\lambda}$ is propagated numerically. 

\begin{algorithm}[H]
\label{alg:var_qdyn}
\caption{Variational dynamics of a superposition of \glspl{FG}}
\KwIn{Initial set of parameters 
$\bm{\lambda}(0) = \left[\bm{\theta}(0), \bm{x}(0), \bm{p}(0), \bm{\gamma}(0)\right]$, 
where
$\bm{\theta}(0)=\left[\theta_{1}(0), \theta_{2}(0), ..., \theta_{N}(0)\right]$, 
$\bm{x}(0)=\left[x_{0}(0), x_{1}(0)..., x_{N}(0)\right]$,
$\bm{p}(0)=\left[p_{0}(0), p_{1}(0)..., p_{N}(0)\right]$,
$\bm{\gamma}(0)=\left[\gamma_{1}(0), \gamma_{2}(0), ..., \gamma_{N}(0)\right]$
}
\KwOut{Trajectories in parameter space $\bm{\lambda}(t) = \left[\bm{\theta}(t), \bm{x}(t), \bm{p}(t), \bm{\gamma}(t)\right]$}
Set
$\bm{\lambda} \gets \bm{\lambda}(0)$ \;
\For{$t = 0$ \KwTo $T-1$}
{
    Compute $\text{Im}(\bm{M}) \gets \text{Im}(\bm{M}(\bm{\lambda}))$ \;
    Compute $\text{Re}(\bm{V}) \gets \text{Re}(\bm{V}(\bm{\lambda}))$ \;
    Solve for $\dot{\bm{\lambda}}$ in $\text{Im}(\bm{M})\dot{\bm{\lambda}}=-\frac{1}{\hbar}\text{Re}(\bm{V})$\;
    Update $\bm{\lambda} \gets \textsc{Integrate}(\bm{\lambda}, \dot{\bm{\lambda}}, \Delta t)$\;
    Store $\bm{\lambda}$\;
}
\Return $\bm{\lambda}(t)$\;
\end{algorithm}

\begin{figure}[H]\centering
		\resizebox{1.0\textwidth}{!}{
\begin{quantikz}[row sep=0.8 cm, column sep=0.1cm]
\lstick{$a_{1}:\ket{0}$}     & \gate{H}     & \ctrl{1} & \qw                  & \qw                  & \qw 
& \qw                  & \qw              & \qw      & \ctrl{1} & \qw                  & \qw                  
& \qw               & \qw                 & \qw               & \qw      & \ctrl{1} & \targ{}  & \ctrl{2} 
& \qw                  & \qw                  & \qw              & \qw                  & \qw              
& \qw      & \ctrl{2} & \qw                 & \qw                  & \qw              & \qw                  
& \qw               & \qw      & \ctrl{2} & \ctrl{1} & \ctrl{2} & \gate{H} & \meter{$\sigma_{z}$} \\
\lstick{$a_{2}:\ket{0}$}     & \qw          & \gate{H} & \ctrl{2}             & \ctrl{4}             & \qw 
& \qw                  & \qw              & \ctrl{2} & \targ{}  & \ctrl{2}             & \ctrl{4}             
& \qw          & \qw                 & \qw               & \ctrl{2} & \gate{H} & \gate{S} & \qw 
& \qw                  & \qw                  & \qw              & \qw                  & \qw              
& \qw      & \qw      & \qw                 & \qw                  & \qw              & \qw                  
& \qw               & \qw      & \qw      & \targ{} & \qw      & \qw & \qw \\
\lstick{$a_{3}:\ket{0}$}     & \qw          & \qw      & \qw                  & \qw                  & \qw 
& \qw                  & \qw              & \qw      & \qw      & \qw                  & \qw                  
& \qw               & \qw                 & \qw               & \qw      & \qw      & \qw      & \gate{H} 
& \ctrl{2}             & \ctrl{3}             & \qw         & \qw                  & \qw              
& \ctrl{2} & \targ{} & \ctrl{2}             & \ctrl{3}             & \qw         & \qw                  
& \qw               & \ctrl{2} & \gate{H} & \qw     & \targ{}  & \qw & \qw \\
\lstick{$g_{n}:\ket{0}^{N}$} & \qwbundle{N} & \qw      & \gate{R_{y_{n}}^{+}} & \qw                  & \qw 
& \ctrl{2}             & \qw         & \gate{H} & \qw      & \gate{R_{y_{n}}^{-}} & \qw                  
& \qw               & \ctrl{2}             & \qw         & \gate{H} & \qw     & \qw       & \qw 
& \qw                  & \qw                  & \qw              & \qw                  & \qw              
& \qw      & \qw     & \qw                  & \qw                  & \qw              & \qw                  
& \qw               & \qw      & \qw      & \qw     & \qw      & \qw & \qw \\
\lstick{$h_{n}:\ket{0}^{N}$} & \qwbundle{N} & \qw      & \qw                  & \qw                  & \qw 
& \qw                  & \qw              & \qw      & \qw      & \qw                  & \qw                  
& \qw               & \qw                  & \qw              & \qw      & \qw     & \qw       & \qw 
& \gate{R_{y_{n}}^{p}} & \qw                  & \qw              & \ctrl{1}             & \qw         
& \gate{H} & \qw     & \gate{R_{y_{n}}^{m}} & \qw                  & \qw              & \ctrl{1}             
& \qw          & \gate{H} & \qw      & \qw     & \qw      & \qw & \qw \\
\lstick{$s:\ket{0}$}         & \qw          & \qw      & \qw                  & \gate[2]{U_{0}^{+}} & \qw 
& \gate[2]{U_{n}^{+}} & \qw              & \qw      & \qw      & \qw                  & \gate[2]{U_{0}^{-}} 
& \qw               & \gate[2]{U_{n}^{-}} & \qw              & \qw      & \qw     & \qw       & \qw 
& \qw                  & \gate[2]{U_{0}^{p}} & \qw              & \gate[2]{U_{n}^{p}} & \qw              
& \qw     & \qw      & \qw                  & \gate[2]{U_{0}^{m}} & \qw              & \gate[2]{U_{n}^{m}} & 
\qw                 & \qw      & \qw      & \qw     & \qw      & \qw & \qw \\
\lstick{$\ket{0}_{m}$}       & \gate[style={dashed}]{S_{\text{I}}}          & \gate[style={dashed}]{D_{\text{I}}}      & \qw                  & \qw                  & \qw 
& \qw                  & \qw & \qw      & \qw      & \qw                  & \qw                  
& \qw  & \qw                  & \qw & \qw      & \qw     & \qw       & \qw 
& \qw                  & \qw                 & \qw  & \qw                  & \qw 
& \qw     & \qw      & \qw                 & \qw                   & \qw & \qw                 
& \qw & \qw       & \qw      & \qw     & \qw      & \qw & \qw    
\end{quantikz}
}
\caption{This circuit approximates $\text{Im}(M_{jk})\approx  \frac{2^{N}\braket{\sigma_{z}}}{\epsilon_{j}\epsilon_{k}}$ for a superposition of $2^{N}$ \glspl{GWP}. It embeds wave function derivatives by finite central difference along variational parameters $\lambda_{j}$ and $\lambda_{k}$. Gates with the superscripts $+$ and $p$ induce positive shifts, while those with superscripts $-$ and $m$ induce negative shifts. Gates with superscripts $+$ and $-$ shift $\lambda_{j}$, while those with superscripts $p$ and $m$ shift $\lambda_{k}$. 
Parameters are not all shifted simultaneously. Computing $\text{Im}(M_{x_{0}\theta_{3}})$ for example, requires shifts in $U_{0}^{+}(\gamma_{0}, x_{0}+\epsilon_{j}, p_{0})$, $U_{0}^{-}(\gamma_{0}, x_{0}-\epsilon_{j}, p_{0})$, $R_{y_{3}}^{p}(\theta_{3}+\epsilon_{k})$ and $R_{y_{3}}^{-}(\theta_{3}-\epsilon_{k})$ only.
Since $\text{Im}(\bm{M})$ is antisymmetric, the circuit only needs to run for unordered pairs of $j$ and $k$.}
\label{fig:Mjk_mgwp_circ}
\end{figure}
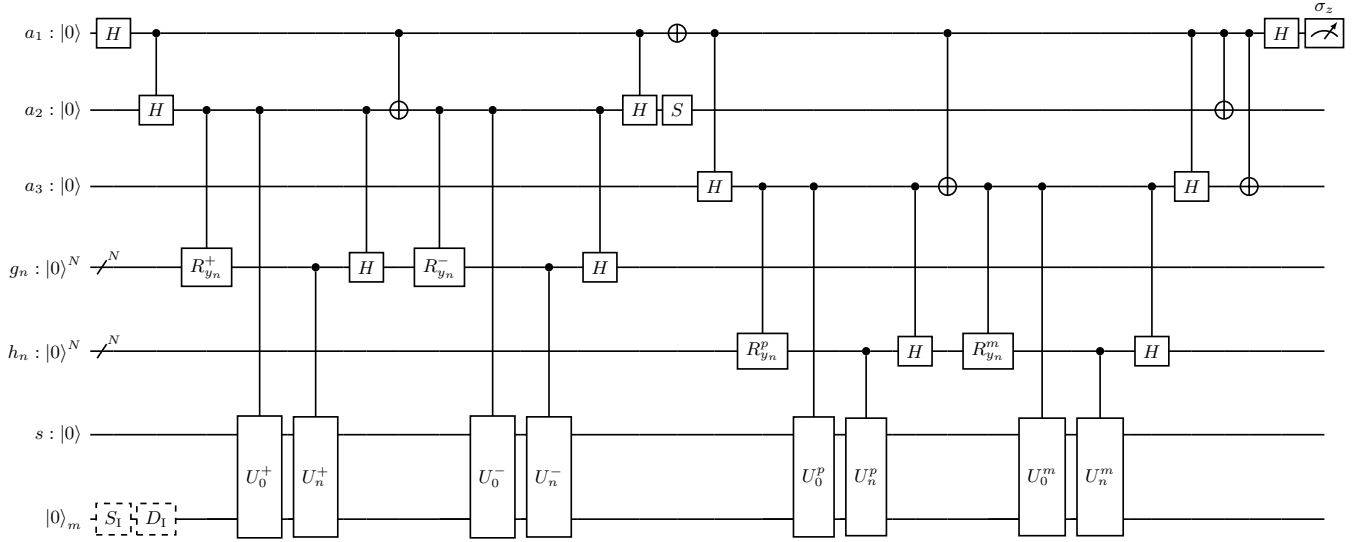

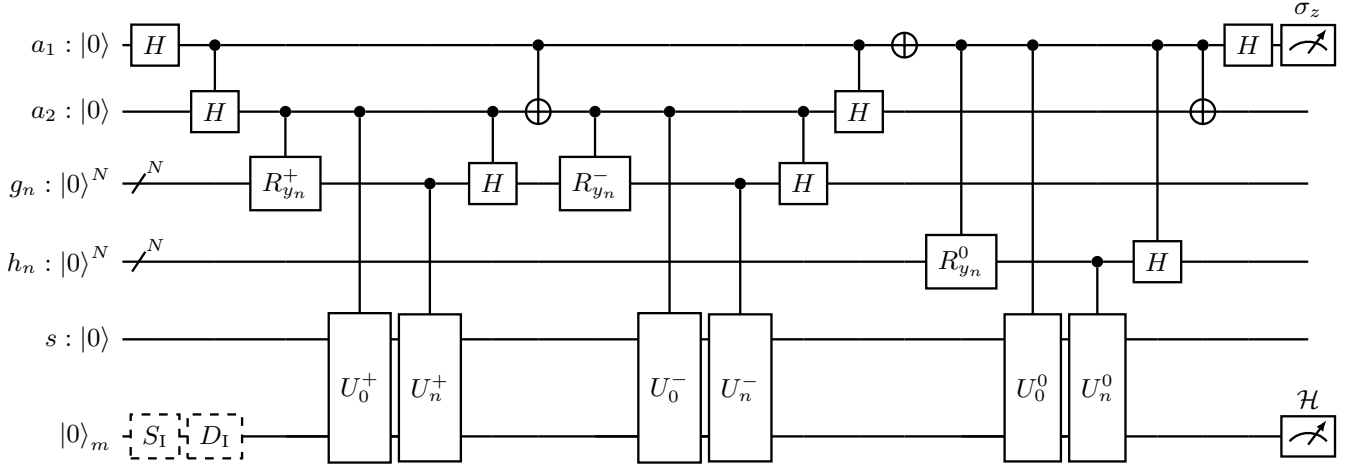
\begin{figure}[H]\centering
		\resizebox{1.0\textwidth}{!}{
\begin{quantikz}[row sep=0.3 cm, column sep=0.1cm]
\lstick{$a_{1}:\ket{0}$} & \gate{H} & \ctrl{1} & \qw & \qw  & \qw & \qw & \ctrl{1} & \qw & \qw & \qw & \qw & \ctrl{1} & \targ{} & \ctrl{3} & \ctrl{4} & \qw & \ctrl{3} & \ctrl{1} & \gate{H} & \meter{$\sigma_{z}$} \\
\lstick{$a_{2}:\ket{0}$} & \qw & \gate{H} & \ctrl{1} & \ctrl{3} & \qw & \ctrl{1} & \targ{} & \ctrl{1} & \ctrl{3} & \qw & \ctrl{1} & \gate{H} & \qw & \qw & \qw & \qw & \qw & \targ{} & \qw & \qw \\
\lstick{$g_{n}:\ket{0}^{N}$} & \qwbundle{N} & \qw & \gate{R_{y_{n}}^{+}} & \qw & \ctrl{2} & \gate{H} & \qw      & \gate{R_{y_{n}}^{-}} & \qw & \ctrl{2} & \gate{H} & \qw & \qw & \qw & \qw & \qw & \qw & \qw & \qw & \qw \\
\lstick{$h_{n}:\ket{0}^{N}$} & \qwbundle{N} & \qw & \qw & \qw & \qw & \qw & \qw & \qw & \qw & \qw & \qw & \qw & \qw & \gate{R_{y_{n}}^{0}} & \qw & \ctrl{1} & \gate{H} & \qw & \qw & \qw \\
\lstick{$s:\ket{0}$} & \qw & \qw & \qw & \gate[2]{U_{0}^{+}} & \gate[2]{U_{n}^{+}} & \qw & \qw & \qw & \gate[2]{U_{0}^{-}} & \gate[2]{U_{n}^{-}} & \qw & \qw & \qw & \qw & \gate[2]{U_{0}^{0}} & \gate[2]{U_{n}^{0}} & \qw & \qw & \qw & \qw \\
\lstick{$\ket{0}_{m}$} & \gate[style={dashed}]{S_{\text{I}}} & \gate[style={dashed}]{D_{\text{I}}} & \qw & \qw & \qw & \qw & \qw & \qw & \qw & \qw & \qw & \qw & \qw & \qw & \qw & \qw & \qw & \qw & \qw & \meter{$\mathcal{H}$} 
\end{quantikz}
}
\caption{This circuit approximates $\text{Re}(V_{j}) \approx \frac{2^{N} \mathbb{E}\left[\left(-1\right)^{f}\mathcal{H}_{m}\right]}{\epsilon_{j}}$ for a superposition of $2^{N}$ \glspl{GWP}, where $f=0$ if $a_{1}$  collapses to $\ket{0}$ or $f=1$ if it collapses to $\ket{1}$, and $\mathcal{H}_{m}$ is the m\textsuperscript{th} measurement of $\mathcal{\hat{H}}$. It embeds the derivative of the  wave function by finite central difference along variational parameter $\lambda_{j}$. Like Fig.~\ref{fig:Mjk_mgwp_circ}, gates with the superscript $+$ induce positive shifts, and those with the superscript $-$ induce negative shifts.  
Parameters are not all shifted simultaneously. Computing $\text{Re}(V_{\gamma_{5}})$ for example, requires shifts in $U_{5}^{+}(\gamma_{5}+\epsilon_{j}, x_{5}, p_{5})$ and $U_{5}^{-}(\gamma_{5}-\epsilon_{j}, x_{5}, p_{5})$  only. Gates with superscript $0$ do not shift the parameters.}
\label{fig:Vj_mgwp_circ}
\end{figure}

We employ the same approach as before, namely approximating $M_{jk}(\bm{\lambda})$ and $V_{j}(\bm{\lambda})$ by finite central difference. In the case of $M_{jk}(\bm{\lambda})$, we embed $\ket{\Psi(\lambda_{j} + \epsilon_{j})}$, $\ket{\Psi(\lambda_{j} - \epsilon_{j})}$, $\ket{\Psi(\lambda_{k} + \epsilon_{k})}$ and $\ket{\Psi(\lambda_{k} - \epsilon_{k})}$ to different computational bases of the ancilla system and apply controlled-Hadamard and CNOT gates so that statistical measurement average approximates 
$\text{Im}(\langle\Psi(\lambda_{j} + \epsilon_{j}) - \Psi(\lambda_{j} - \epsilon_{j})\mid \Psi(\lambda_{k} + \epsilon_{k}) - \Psi(\lambda_{k} - \epsilon_{k})\rangle)$. The same is done for $V_{j}(\bm{\lambda})$, where we embed $\ket{\Psi(\lambda_{j} + \epsilon_{j})}$, $\ket{\Psi(\lambda_{j} - \epsilon_{j})}$ and $\ket{\Psi(\bm{\lambda})}$. The measurement averages are proportional to $\text{Re}(\langle\Psi(\lambda_{j} + \epsilon_{j}) - \Psi(\lambda_{j} - \epsilon_{j})\mid \hat{\mathcal{H}} \mid \Psi(\bm{\lambda})\rangle)$. The circuit approximating $\text{Im}(M_{jk}(\bm{\lambda}))$ is depicted in Fig.~\ref{fig:Mjk_mgwp_circ} and the one approximating  $\text{Re}(V_{j}(\bm{\lambda}))$ is depicted in Fig.~\ref{fig:Vj_mgwp_circ}.

\section{Numerical simulations}

\subsection{Computational details}
\label{sec:comput-details}

This section presents the results of numerical simulations of the variational dynamics algorithms of a single and a superpositions of \glspl{FG} presented in Sec.~\ref{sec:var_dyn_single} and Sec.~\ref{sec:var_dyn_superpos}. Three \glspl{PES} models were tested: quadratic, Morse and double-well potentials. For all cases, we assess the algorithm's abilities to simulate the autocorrelation function $\mu(t)=\braket{\Psi(0)|\Psi(t)}$, given our ongoing interest in vibronic spectroscopy.  Vibronic spectra can be obtained directly from $\mu(t)$ via Fourier transform and $\mu(t)$ can be obtained from wavepacket dynamics. 

We simulated wavepacket dynamics following the instantaneous excitation from an initial quadratic potential $V_{\text{I}}(r) = \frac{1}{2}\omega_{\text{I}}^{2}(r-r_{\text{I}})^{2}$ to a final potential $V_{\text{F}}(r)$. Choosing a quadratic form for $V_{\text{I}}(r)$ is appropriate given that many spectroscopic processes probe molecules near their equilibrium where the leading Taylor expansion term is quadratic.  The initial wave function is thus the ground state of $V_{\text{I}}(r)$ and this setup emulates vibronic processes invoked in dipole-allowed optical transitions between pairs of electronic states. 

The variational parameters $\bm{\theta}$, $\bm{\gamma}$ and $\bm{p}$ were initially set to $\bm{0}$. Although ancillary Gaussians are not initially occupied, their initial positions, set by $\bm{x}$, need to be carefully chosen. For example, the guiding and ancillary Gaussians should not strongly overlap as it introduces redundancies in the space of variational parameters causing the columns and rows of $\bm{M}$ to be linearly dependent. Thus, we instantiated $x_{0}=0$ and $x_{n}=-2^{n}\Delta x$. The set $\{x_0, x_1, ...,x_N, \}$ is sum-distinct, meaning that every one of its subset sum is unique. Each subset sum coincides with the center of one of the Gaussians. The $\Delta x$ parameter is the difference between centers of nearest neighbor Gaussians. A large $\Delta x$ spreads Gaussians apart, while a small value brings them closer together. The negative sign ensures that ancillary \glspl{FG} are present on both sides of the guiding Gaussian. For each simulation, we experimented with different values of $\Delta x$ and chose the ones that performed best. While a more systematic way of selecting the initial parameters may yield improved performance, we leave this refinement to future investigations. 

We used the split-operator method~\cite{kosloff1988time} to perform numerically exact quantum dynamics. We used a time step of 0.05~\text{fs} and a spatial grid of 10001 points uniformly distributed over the interval $[-500, 500]$. Eqs.~\eqref{eq:MVP} and \eqref{eq:KSVP} were solved numerically using the LSODA method~\cite{petzold1983automatic} as implemented in SciPy~\cite{2020SciPy-NMeth} with absolute and relative tolerances of $10^{-9}$ and $10^{-6}$. Although the numerical integrator uses an adaptive time step, $\mu(t)$ was constructed at every interval of 0.05~\text{fs}, unless otherwise noted. The \glspl{VQA} were emulated classically by representing the qumode and the \gls{CV} gates in a harmonic oscillator eigenbasis truncated to the 80 lowest states. We used $\epsilon = 0.001$ for all variational parameter shifts to run the circuits computing the real and imaginary parts of $\bm{M}$ and $\bm{V}$. One of the short-comings of our approach, also present in other variational multi-Gaussian methods,~\cite{sawada1985strategy, worth2020gaussian} is that $\text{Im}(\bm{M})$ is often nearly singular. To mitigate this issue, we used Tikhonov regularization~\cite{tikhonov1943stability} with a regularization factor of $10^{-8}$. For all models, we verified that the results were insensitive to the regularization factor. We experimented with regularization factors of $10^{-10}$, $10^{-9}$, $10^{-8}$ and $10^{-7}$ and obtained nearly identical dynamics in all models. All model parameters were converted to mass-weighted atomic units, which were used for subsequent dynamical simulations. 

\subsection{Single \gls{FG} dynamics: \ce{SO2} spectrum from physical quantum hardware}

A single \gls{FG} is a reasonable approximation when $V_{\text{I}}(r)$ and $V_{\text{F}}(r)$ are quadratic potentials with similar frequencies. This is the case in the dynamics underlying the photoelectron spectrum of \ce{SO2}. As a first step towards a full experimental realization of the proposed algorithms, we compute $\mu(t)$ of the photoexcited dynamics of \ce{SO2} on physical quantum hardware.

Doing so with the setup considered herein requires a 1D model of \ce{SO2}~\cite{macdonell2023predicting}. Although \ce{SO2} has three vibrational degrees of freedom, one of them \,---\, the bending mode \,---\, happens to be far more displaced than the other two and thereby dominates the short time dynamics. Its 1D quadratic potential was constructed as follows. First, we computed the Hessians of \ce{SO2} and \ce{SO2+} at their minima using the Firefly software package~\cite{FFly} with the B3LYP/6-311G(d,p)~\cite{becke1993new, lee1988development, krishnan1980self} level of theory. The restricted open-shell formalism was used for \ce{SO2+}. Then, the relative displacement $r_{\text{I}}$ between the harmonic potentials $V_{\text{I}}(r)$ and $V_{\text{F}}(r)$ was computed $r_{\text{I}} = \bm{\Lambda}_{b}\bm{M_{a}^{\frac{1}{2}}\left(X_{\text{I}}-X_{\text{F}}\right)}$, where $\bm{\Lambda}_{b}$ is the eigenvector of the mass-weighted Hessian of \ce{SO2+} corresponding to the bending mode, $\bm{M_{a}}$ is a square diagonal matrix containing the atomic masses, and $X_{\text{I}}$ and $X_{\text{F}}$ are the Cartesian coordinates of the minima of \ce{SO2} and \ce{SO2+}, respectively. The frequencies of the bending modes $\omega_{\text{I}}$ and $\omega_{\text{F}}$ where taken from the Hessian calculations. The parameters in mass-weighted atomic units are $\omega_{\text{I}}=0.00231$, $\omega_{\text{F}}=0.00185$ and $r_{\text{I}} = -54.5$.

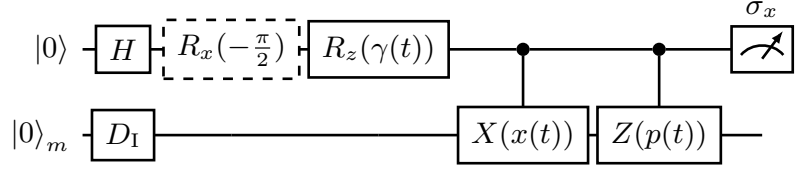
\begin{figure}[H]\centering
		\resizebox{0.6\textwidth}{!}{
\begin{quantikz}[row sep=0.3 cm, column sep=0.1cm]
\lstick{$\ket{0}$} & \gate{H} & \gate[style={dashed}]{R_{x}(-\frac{\pi}{2})} & \gate{R_{z}(\gamma(t))} & \ctrl{1} &\ctrl{1} & \meter{$\sigma_{x}$} \\
\lstick{$\ket{0}_{m}$} & \gate{D_{\text{I}}} & \qw & \qw & \gate{X(x(t))} & \gate{Z(p(t))} & \qw
\end{quantikz}
}
\caption{Circuit computing $\mu(t)$ for a single \gls{FG} defined by the variational parameters $\left[x_{t}, p_{t}, \gamma_{t}\right]$. Running the circuit while omitting the dashed gate outputs $\text{Re}\left(\mu(t)\right)$; including it outputs $\text{Im}\left(\mu(t)\right)$.}
\label{fig:FG_mu_circ}
\end{figure}

The \gls{FG} was propagated by executing the \gls{VQA} on a classical computer for a total duration of 400~\text{fs}. At every 0.1~\text{fs}, the resulting variational parameters $x(t)$, $p(t)$ and $\gamma(t)$ were used as input in the physical circuit depicted in~\Cref{fig:FG_mu_circ}, evaluating $\mu(t)$.

In the following, we provide a brief description of the hardware used for the experiment and discuss the implementation of the autocorrelation circuit on our platform.

\begin{figure}[htb]
\centering

\subfloat[]{%
    \includegraphics[width=0.45\textwidth]{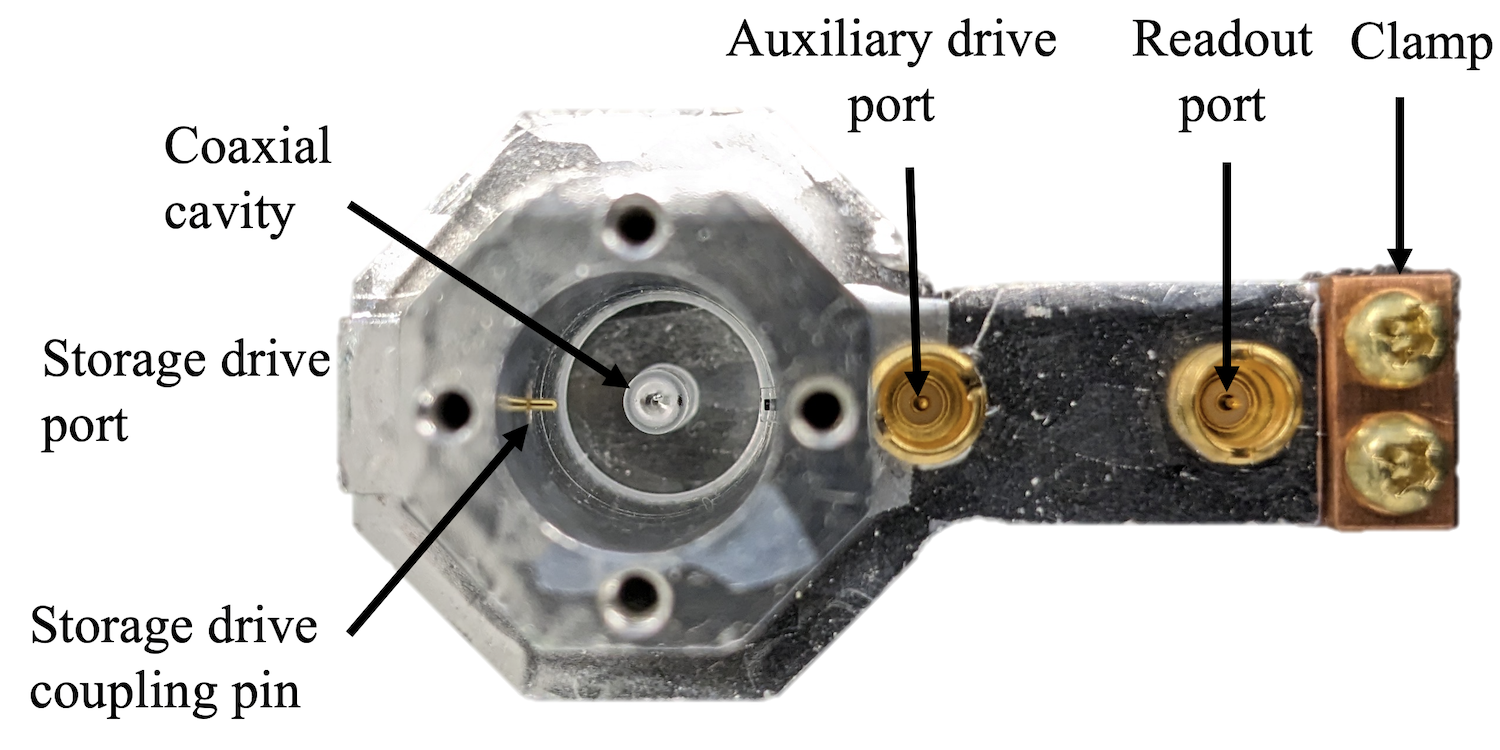}
}\hfill
\subfloat[]{%
    \includegraphics[width=0.25\textwidth]{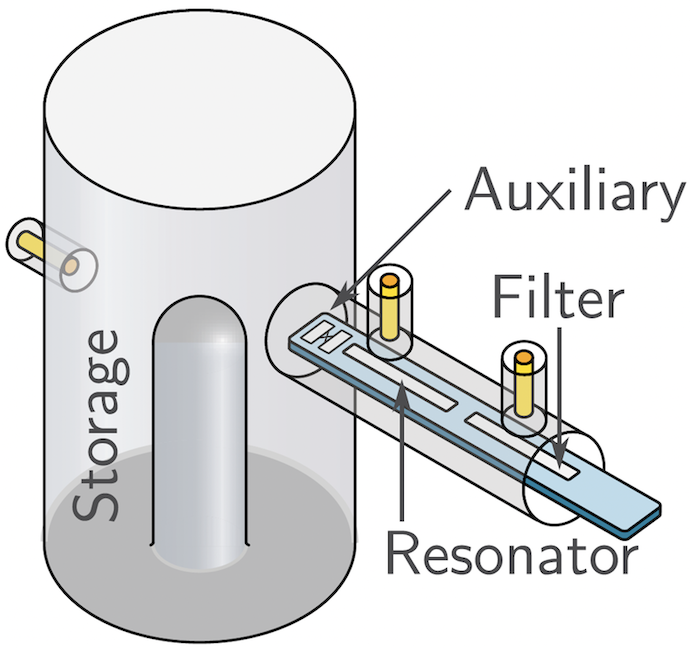}
}\hfill
\subfloat[]{%
    \includegraphics[width=0.15\textwidth]{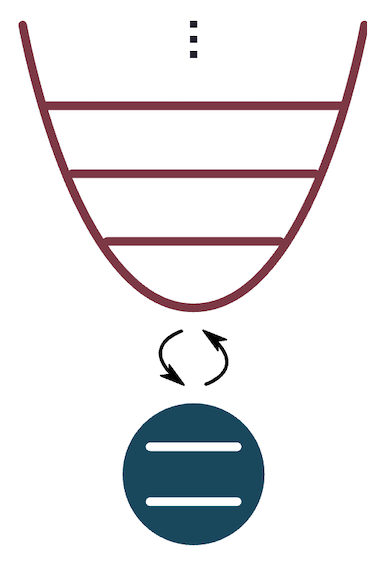}
}

\captionsetup{justification=raggedright}
\caption{(a) Photo of the hardware assembly. (b) Schematic of hardware architecture. A $3$-dimensional coaxial single-mode cavity dispersively coupled to a transmon ancilla connected to a readout resonator. (c) The storage mode can be considered as harmonic oscillator interacting with a two-level system, the auxiliary transmon.}
\label{fig:HW_figs}
\end{figure}

Our setup is a superconducting oscillator-transmon quantum hardware composed of a single-mode superconducting microwave cavity dispersively coupled to an auxiliary transmon, as shown in~\Cref{fig:HW_figs}. The cavity shown in~\Cref{fig:HW_figs}(a) is machined from a block of high purity aluminum, held in a three-layer cryoperm-niobium-cryoperm magnetic shield and hosts a long-lived bosonic mode that serves as a storage mode. The chip, comprising the auxiliary transmon and the readout resonator, is placed in a waveguide with a copper clamp. The readout, the auxiliary, and the storage drive ports are used to drive the system.

Since the auxiliary transmon is weakly dispersively coupled to both the storage and resonator, the Jaynes-Cummings Hamiltonian 
can be expressed to second-order in perturbation theory as~\cite{blais2021circuit, lachance2024autonomous}
\begin{equation}    \hat{H}_{\text{disp}}/\hbar=\omega_s\hat{a}^\dagger\hat{a}+\omega_q\hat{b}^\dagger\hat{b}+\omega_r\hat{c}^\dagger\hat{c}+ 2\chi_{sq}\hat{a}^\dagger\hat{a}\hat{b}^\dagger\hat{b}+2\chi_{qr}\hat{b}^\dagger\hat{b}\hat{c}^\dagger\hat{c}+\cdots,
\end{equation}
where $\omega_{s}$, $\omega_{q}$ and $\omega_{r}$ are the dressed frequencies of the storage, the qubit and the resonator modes. Moreover, $\hat{a}$, $\hat{b}$ and $\hat{c}$ are the annihilation operators corresponding to the dressed modes and $\chi_{sq}$ and $\chi_{qr}$ are storage-qubit and qubit-resonator linear dispersive shift, respectively. The dispersive coupling provides the entangling interaction underlying controlled operations between the qubit and qumode. More details about the hardware can be found in Ref.~\cite{lachance2024autonomous}. 

The oscillator–transmon architecture employed in this work was originally developed to realize fault-tolerant Gottesman--Kitaev--Preskill logical qubits~\cite{lachance2024autonomous}. In addition to its initial purpose, it naturally supports \gls{CV-DV} quantum computation by treating the storage cavity as a qumode and the transmon as a qubit. 
Universal control of a bosonic mode requires an entangling operation between the linear cavity mode and the nonlinear ancilla. A key native primitive is an \gls{ECD} gate, which implements a cavity displacement with programmable amplitude and phase conditioned on the transmon state through the storage-transmon cross-Kerr interaction, $\chi_{sq}$~\cite{eickbusch2022fast,lachance2024autonomous}. Beam-splitter interaction, arbitrary single-qubit rotations and ECD gates form a universal gate set for \gls{CV-DV} quantum computation.

We attempted to perform nuclear dynamics simulations by computing $\text{Re}\left(\bf{M}\right)$ and $\text{Im}\left(\bf{V}\right)$ matrix elements on our quantum hardware. However, we were unsuccessful because several matrix elements were of the order of $10^{-7}$; such magnitudes require very high precision when performing $\langle\sigma_x\rangle$ and $\langle\mathcal{H}\rangle$ measurements and are very challenging given that current readout errors are of the order of $10^{-2}$. Due to this restriction, we were limited to using classically obtained variational parameters to compute the autocorrelation function on the quantum hardware. 
We transpiled the autocorrelation circuit in~\Cref{fig:FG_mu_circ} into a sequence of the hardware-native gate set, composed of cavity displacement, \gls{ECD}~\cite{eickbusch2022fast, lachance2024autonomous} and auxiliary rotation gates. The real and imaginary parts of the autocorrelation function $\mu(t)$ were obtained by executing the corresponding circuits with 1000 shots.

\begin{figure}[H]
	\centering
	\includegraphics[trim=100 0 80 30, clip, width=1.0\textwidth]{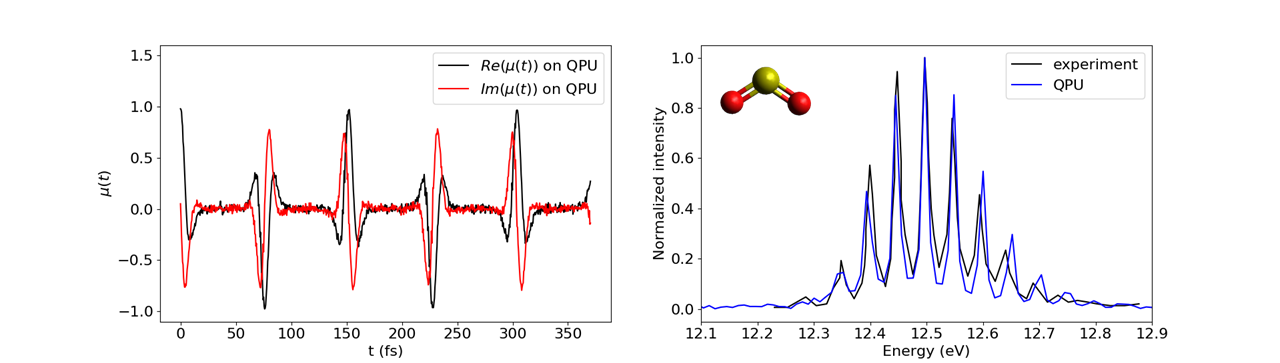}
	\caption{(Left) Real and imaginary parts of the autocorrelation function evaluated using physical quantum hardware. (Right) Experimental and quantum hardware-derived photoelectron spectra. The blue line is the spectrum obtained from running the circuits in~\Cref{fig:Mjk_gauss_circuit,fig:Vj_gauss_circuit} on classical hardware, and from~\Cref{fig:FG_mu_circ} on quantum hardware. The black line is the experimental spectrum~\cite{holland1994experimental}.}
	\label{fig:Nqubits_0_so2}
\end{figure}

The left side of~\Cref{fig:Nqubits_0_so2} features $\mu(t)$ obtained from experimental quantum hardware. On the right is the photoelectron spectrum obtained from classical Fourier transform of $\mu(t)$ and subject to a 12.3~\text{eV} horizontal translation to make it overlap with the experimental spectrum~\cite{holland1994experimental}. As evident from~\Cref{fig:Nqubits_0_so2}, the \gls{VQA} performs reasonably well; it correctly reproduces both spacings and relative intensities of vibronic peaks. 

\subsection{Thawed Gaussian dynamics}
\label{sec:undisplHarmPot}

The \glspl{VQA} of a single and a superposition of \glspl{FG} dynamics were implemented to simulate wavepacket dynamics on the quadratic potential $V_{\text{F}}(r) = \frac{1}{2}\omega_{\text{F}}^{2}r^{2}$ with model parameters $r_{\text{I}} = 0.0$, $\omega_{\text{I}} = 9.557\times 10^{-3}$ and $\omega_{\text{F}} = 3.3515\times 10^{-3}$. The numerical value chosen for $\omega_{\text{F}}$ is one of the two harmonic frequencies belonging to a 2D-LVC model of photoionized butatriene~\cite{ryabinkin2014we}. It was selected to ensure the energetics were in the range of typical nuclear vibrations. When $\omega_{\text{I}} \neq \omega_{\text{F}}$, the exact solution to the \gls{TDSE} is a single thawed Gaussian whose width varies periodically in time. The objective here is to test whether a superposition of \glspl{FG} evolving according to the \gls{VQA} can simulate thawed Gaussian dynamics. For this reason, we set $\omega_{\text{I}}$ to be approximately three times larger than $\omega_{\text{F}}$. 

\begin{figure}[H]
	\centering
	\includegraphics[trim=120 30 90 20, clip=true, width=1.0\textwidth]{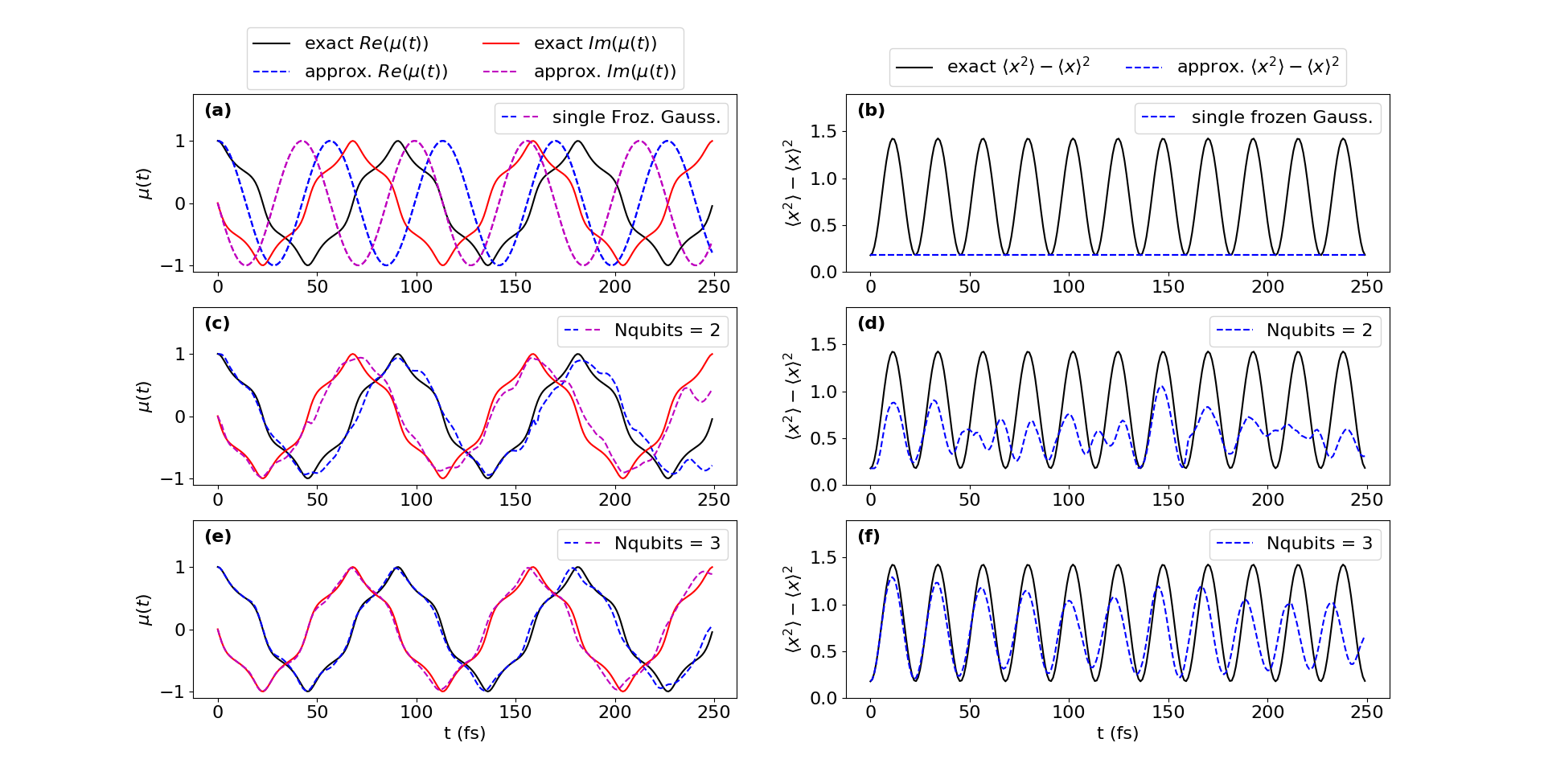}
	\caption{Panels (a), (c) and (e) compare the real and imaginary parts of $\mu(t)$ of exact thawed Gaussian dynamics with those obtained from the \glspl{VQA} of a single, and superpositions of 4 and 8 \glspl{FG}, respectively. Panels (b), (d) and (f) compare the width of the exact dynamics with those of a single, and superpositions of 4 and 8 \glspl{FG}, respectively.}
	\label{fig:nondispl_harm_buta}
\end{figure}

To gauge the performance, we compare the real and imaginary parts of $\mu(t)$ and the variance of position $\braket{\hat{x}^{2}}-\braket{\hat{x}}^2$ between exact and approximative dynamics in~\Cref{fig:nondispl_harm_buta}. For the variance, we use $\hat{x} = \sqrt{\frac{1}{2}}\left(\hat{a}^{\dagger} + \hat{a}\right)$, omitting the harmonic frequency. The real and imaginary parts of $\mu(t)$ of the superposition of Gaussians can be obtained from the quantum circuit depicted in Fig.~\ref{fig:multi_FG_mu_circ}.

\begin{figure}[H]\centering
		\resizebox{0.9\textwidth}{!}{
\begin{quantikz}[row sep=0.3 cm, column sep=0.1cm]
\lstick{$a_{1}:\ket{0}$} & \gate[style={dashed}]{R_{x}(-\frac{\pi}{2})} & \gate{H} & \ctrl{1} & \ctrl{2} & \qw & \ctrl{1} & \gate{H} & \meter{$\sigma_{z}$} \\
\lstick{$g_{n}:\ket{0}^{N}$} & \qwbundle{N} & \qw & \gate{R_{y_{n}}(\theta_{n}(t))} & \qw & \ctrl{1} & \gate{H} & \qw & \qw \\
\lstick{$s:\ket{0}$} & \qw & \qw & \qw & \gate[2]{U(\gamma_{0}(t),x_{0}(t), p_{0}(t))} & \gate[2]{U(\gamma_{n}(t),x_{n}(t), p_{n}(t))} & \qw & \qw & \qw \\
\lstick{$\ket{0}_{m}$} & \gate[style={dashed}]{S_{I}} & \gate[style={dashed}]{D_{I}} & \qw & \qw & \qw & \qw & \qw & \qw 
\end{quantikz}
}
\caption{Circuit computing $\mu(t)$ for a superposition of \glspl{FG}. Running the circuit omitting the dashed $R_{x}(-\frac{\pi}{2})$ gate outputs measurements where the average $\braket{\hat{\sigma}_{z}} = Re\left(\mu(t)\right)$; including it outputs measurements where the average $\braket{\hat{\sigma}_{z}} = Im\left(\mu(t)\right)$.}
\label{fig:multi_FG_mu_circ}
\end{figure}

As shown in panels (a) and (b) of~\Cref{fig:nondispl_harm_buta}, unsurprisingly, a single \gls{FG} fails to capture the exact wavepacket properties. A superposition of four \glspl{FG} simulates relatively well $\mu(t)$, as illustrated in panel (c), but performs somewhat poorly in simulating width dynamics as evident in panel (d). While there is apparent oscillatory behavior in its width more or less resonant with the exact solution, the amplitudes are comparatively attenuated. Panels (e) and (f) show that by doubling the number of Gaussians the approximate wavepacket produces a more accurate $\mu(t)$ and substantially better width dynamics with amplitudes that are closer to the exact solution.

\subsection{Morse potential: nuclear dynamics on the {$B^{1}\Sigma_{u}^{+}$} state of \ce{H2}}

In this section, we compare the variational wavepacket dynamics with the numerically exact solution on the potential energy surface of the $B^{1}\Sigma_{u}^{+}$ state of \ce{H2}, its lowest stable excited state~\cite{kol1966potential}. It can be fitted to a Morse potential $V_{\text{F}}(r) = D_{e}\left(e^{-2ar}-2e^{-ar}\right)$ with the mass-weighted parameters $D_{e} = 1.7436 \times 10^{-1}$ and $a = 3.379 \times 10^{-2}$.  Specifically, we compare their $\mu(t)$ and the extend to which their wavepackets become non-Gaussian. Since a wavepacket that is initially Gaussian does not remain so when it evolves on a Morse potential, an appropriate test is to monitor and compare how their skewness $\left\langle\frac{\hat{x}-\braket{\hat{x}}}{\sqrt{\braket{x^2}-\braket{x}^2}}\right\rangle^{3}$ and kurtosis $\left\langle\frac{\hat{x}-\braket{\hat{x}}}{\sqrt{\braket{x^2}-\braket{x}^2}}\right\rangle^{4}$, both serving as metrics of the non-Gaussianity of the nuclear density, change in time.

\begin{figure}[H]
	\centering
	\includegraphics[trim=120 40 130 20, clip, width=1.0\textwidth]{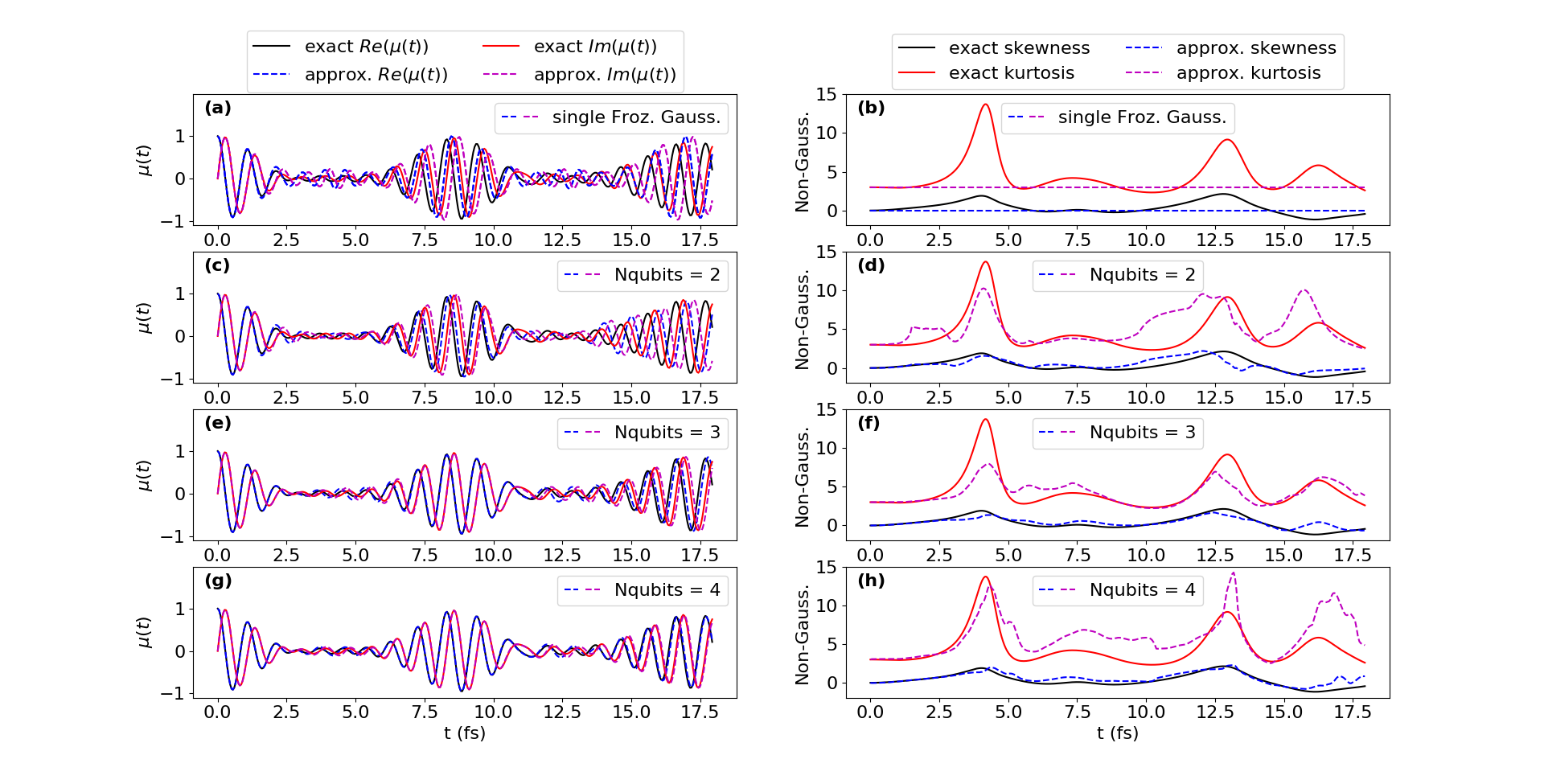}
	\caption{Panels (a), (c), (e) and (g) compare the real and imaginary parts of $\mu(t)$ of exact dynamics on a Morse potential with those obtained variationally with wavepackets composed of different numbers of \glspl{FG}. Panels (b), (d), (f) and (h) compare the time evolutions of the skewness and kurtosis of the nuclear density between exact dynamics and variational schemes with different numbers of \glspl{FG}.}
	\label{fig:Nqbits_morse_h2}
\end{figure}

The initial conditions are defined by $\omega_{\text{I}} = 9.975\times 10^{-3}$ and $r_{\text{I}} = 17.0$. As shown in \Cref{fig:Nqbits_morse_h2} panels (a) and (b), a single \gls{FG} can only describe very short-time dynamics, up to at most 2~\text{fs}; beyond that it is no longer a valid approximation. Panels (c), (e) and (g) show that as the number of Gaussians in the expansion increases, $\mu(t)$ becomes increasingly more accurate. With 16 \glspl{FG}, there is very little distinction between the approximation and the exact solution. Panels (d), (f) and (h) also show that the \gls{VQA} is able to roughly replicate the time evolution of the skewness and kurtosis of the exact wavepacket. While there is a noticable improvement when doubling the number of Gaussians from 4 to 8, there does not seem to be a significant improvement when doubling from 8 to 16 Gaussians, suggesting perhaps that convergence is reached somewhere between 8 and 16 Gaussians.

\subsection{Double-well potential: inversion dynamics of \ce{NH3}}

Quantum dynamics often involves bifurcation. This is when a wavepacket splits into two or more distinct parts that can evolve independently and later interfere coherently. Simulating bifurcation is challenging because it requires getting the right amount of probability density into several places at once and capturing the relative phases between the spatially separated regions correctly. That being said, a more stringent test for the proposed \gls{VQA} is its ability to simulate wavepacket bifurcation. 

We ran the methods on a 1D model of \ce{NH3} and simulated its inversion dynamics. The model is a double-well potential $V_{\text{F}}(r) = \frac{1}{2}\omega_{\text{F}}^{2}r^2 + N_{b}e^{-\zeta r^2}$ with mass-weighted parameters $r_{\text{I}}= 0.0$, $w_{\text{I}} = w_{\text{F}} = 4.03\times 10^{-3}$, $N_{b} = 5.684\times 10^{-2}$ and $\zeta =2.9\times 10^{-4}$~\cite{swalen1962potential}. We chose to place the initial wavepacket directly on the Gaussian barrier. By visualizing the exact numerical dynamics, we see that the ensuing evolution, within the considered total time span, undergoes four phases. In the first, between 0~\text{fs} and 10~\text{fs}, the wavepacket remains Gaussian and widens. In the second, between 10~\text{fs} and 26~\text{fs}, the wavepacket bifurcates into two densities that evolve locally within their respective wells. In the third phase, starting at 26~\text{fs}, the two densities interfere constructively at the barrier and there is a partial recurrence lasting until 46~\text{fs}. After that, the nuclear dynamics become irregular and highly delocalized within the two wells. 

\begin{figure}[H]
	\centering
	\includegraphics[trim=120 30 80 0, clip, width=1.0\textwidth]{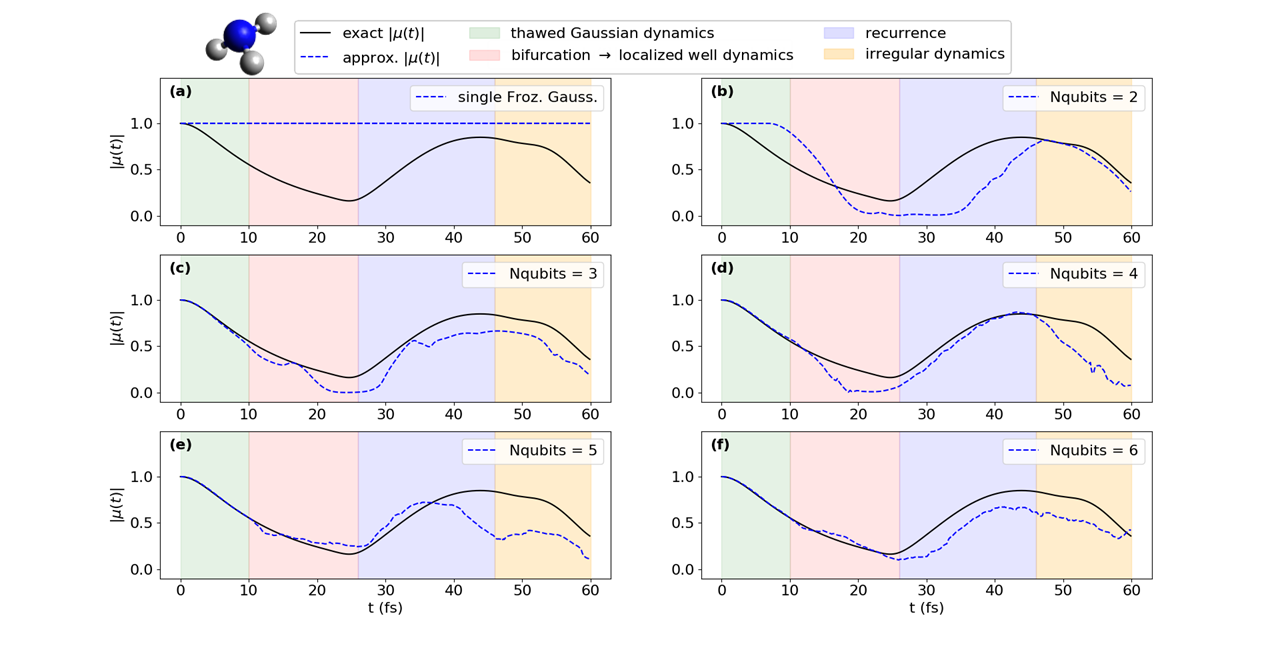}
	\caption{Panels (a), (b), (c), (d), (e) and (f) compare numerically exact $\left|\mu(t)\right|$ with ones obtained from approximate \gls{VQA} dynamics using 1, 4, 8, 16, 32 and 64 \glspl{FG}, respectively. }
	\label{fig:Nqbits_0_double_well}
\end{figure}

\Cref{fig:Nqbits_0_double_well} compares $\left|\mu(t)\right|$ of the exact dynamics with the \gls{VQA} with up to 64 Gaussians. We report here the absolute value because the real and imaginary parts are highly oscillatory and their graphical rendering is cluttered and visually difficult to parse by eye. As expected, single \gls{FG} dynamics, depicted in panel (a), completely fail: the approximative wavepacket remains in its initial position and shape, only picking up a global phase. As shown in panel (b), there is some improvement in using 4 \glspl{FG} despite their inability to correctly capture the proper rate of bifurcation as it is initially too slow before becoming too fast. Whereas the exact solution shows some remnants of nuclear density overlapping with the initial state during localized well dynamics, the approximative solution decays almost completely. Furthermore, the recurrence occurs much later than in the exact dynamics. Simulations with 8 and 16 \glspl{FG}, depicted in panels (c) and (d), improve the bifurcation rate significantly. However their nuclear densities still do not overlap with the initial state during localized well dynamics. Nonetheless, their ability to simulate the recurrence is decent, and the use of 16 \glspl{FG} is superior to 8 \glspl{FG} in reproducing the exact amplitudes and phases. Panels (e) and (f) contain $\mu(t)$ for 32 and 64 \glspl{FG}. They not only simulate the bifurcation rate nearly exactly, but they also capture the correct overlap with the initial state during localized well dynamics. Their recurrence dynamics are moderately good, although slightly inferior to the 16 \glspl{FG} case. Their ability to simulate the irregular dynamics is only moderate, as they struggle to capture the exact amplitudes. 

\subsection{Current limitations}
\label{sec:current-limitations}

In this section, we highlight three limitations of the algorithm: the extremely high cost of stochastic circuit sampling, the lower expressivity of our ansatz compared to \gls{GWP} methods with free variational parameters, and the effect of the initial conditions can affect time steps. 

\subsubsection{Sampling overhead}
A common bottleneck for \glspl{VQA}~\cite{cerezo2021variational,scriva2024challenges} is their potentially high sampling overhead. Our \gls{VQA} has that same problem. Specifically, since matrix elements of $\bm{M}$ and $\bm{V}$ are expectation values approximated through a finite number of measurements, in order for the algorithm to produce accurate dynamics, the statistical averages must closely approximate the circuit's true expectation values. When estimating an expectation value by averaging $N_{shots}$, the standard error on the mean $\varsigma_{x}$ is given by $\varsigma_{x} = \frac{\varsigma}{\sqrt{N_{shots}}}$, where $\varsigma^2$ is the per-shot variance. Since $\bm{M}$ matrix elements correspond to expectations values of $\sigma_{z}$, when $\lvert\left\langle\sigma_{z}\right\rangle\rvert \ll 1$, we obtain $\varsigma^2 \approx 1$. Hence the $N_{shots}$ required to reach a certain level of accuracy as defined by $\varsigma_{x}$ is $N_{shots} \approx \frac{1}{\varsigma_{x}^2}$. 
In the prior case of \ce{SO2}, we evaluated that $\lvert\left\langle\sigma_{z}\right\rangle\rvert \sim 10^{-7}$. Assuming the desirable relative error to be in the surroundings of $1\%$, this would coincide with $\varsigma_{x}^2 \sim 10^{-9}$, thereby requiring $N_{shots} \sim 10^{18}$ for a single matrix element. The astronomical number of shots required makes even the single \gls{FG} variant out of reach for computation on physical quantum hardware. 
The smallness of $\lvert\left\langle\sigma_{z}\right\rangle\rvert$ stems from the finite-difference scheme requiring dividing two small numbers $\left\langle\sigma_{z}\right\rangle$ and $\epsilon_j\epsilon_k$ to compute the matrix elements. Consequently, in order for the method to be applicable on quantum hardware, it becomes necessary to derive some alternative scheme for calculating the matrix elements directly and reduce the sampling overhead.

\subsubsection{Expressivity of the ansatz}

In the context of parameterized quantum circuits, expressivity refers to the range of quantum states an ansatz can produce as a function of its parameters~\cite{sim2019expressibility, ragone2024lie}. Loosely speaking, an ansatz is said to have high expressivity if, by changing its parameters, it can generate diverse quantum states throughout the Hilbert space. Conversely, an ansatz has low expressivity if its parameters can only generate states within a restricted subset of the Hilbert space, so that target states lying outside this subset cannot be approximated regardless of how the parameters are tuned. Our ansatz encodes $2^N$ Gaussians at the cost of $O(N)$ variational parameters, but it is far less expressive than those of classical methods that encode $2^N$ Gaussians using $O(2^N)$ free  variational parameters. In other words, certain Gaussian superpositions represented by Eq.~\ref{eq:mgwf_classical} cannot be represented by Eq.~\ref{eq:mgwf_qregist_ansatz}.

To demonstrate that our ansatz possesses lower expressivity than classical ones, we consider the class of arbitrary superpositions of $2^{N}$ Gaussians, where $N$ is fixed. 
Considering only the position degree of freedom for now, a classical superposition of $2^{N}$ Gaussians corresponds to a point of $\mathbb{R}^{2^{N}}$ (one real position per Gaussian), and is not confined to any proper subspace. On the other hand, in our ansatz, the position of the k$^{\text{th}}$ Gaussian $q_{k}$ is parameterized by $\bm{x}$ where $q_{k} = x_{0} + \sum_{n=1}^{N} \text{bin}(k)_{n} x_{n}$. Hence, a superposition of $2^{N}$ Gaussians built using our ansatz corresponds to a point in a $N+1$ dimensional subspace of $\mathbb{R}^{2^{N}}$. Thus, there are points in $\mathbb{R}^{2^{N}}$ that do not belong to that subspace, which leads to the direct implication that some superpositions of $2^{N}$ Gaussians with $2^{N}$ free variational positional parameters cannot be represented by Eq.~\ref{eq:mgwf_qregist_ansatz}. The same argument can be made using the momentum degree of freedom. 

A similar argument can be made by considering instead the quantum amplitudes. In the limit where the overlap between individual Gaussians is negligible, normalization enforces $\sum_{k}^{2^N} \left|c_{k}\right|^{2} = 1$. The amplitudes of the classical ansatz are thus confined to the positive orthant of the unit sphere $S^{2^{N}-1}$. The amplitudes of the quantum ansatz automatically satisfy normalization, since $\sum_{k=1}^{2^{N}}\prod_{n=1}^{N} \left(\left(\cos^2\left(\frac{\theta_{n}}{2}\right)\right)^{\text{bin}(k)_n}\left(\sin^2\left(\frac{\theta_{n}}{2}\right)\right)^{1-\text{bin}(k)_n}\right) = \prod_{n=1}^{N} \left( \cos^{2}\left(\frac{\theta_{n}}{2}\right)+\sin^{2}\left(\frac{\theta_{n}}{2}\right)\right) = 1$. Thus, they corresponds to a $N$-dimensional submanifold within the positive orthant of $S^{2^{N}-1}$. This has the consequence that certain points on the positive orthant do not belong to the submanifold and that some classical superpositions of Gaussians exist whose amplitudes cannot be reproduced by our ansatz.   

\subsubsection{Arbitrariness of the initial conditions}
The second limitation pertains to the initial conditions. As discussed earlier in Sec.~\ref{sec:comput-details}, in our implementation, Gaussians are given initial positions determined by the parameter $\Delta x$, which controls the initial spread between Gaussians. Because the ancilla Gaussians are not initially populated, $\Delta x$ is arbitrary. 
%However, we noticed while performing numerical simulations that 
The value assigned to it affects the time steps. We found that certain values of $\Delta x$ shrank the time step in the time adaptive integrator 
%to evolve $\bm{\lambda}(t)$
and stalled the propagation.  
%depending in the potential, number of ancilla qubits and initial conditions. 
Usually, we were able to sidestep this issue by tinkering with the value of $\Delta x$ until the propagation reached a manageable time step. 
%In terms of the accuracy, the problem was less severe. We found from the numerics that a very large $\Delta x$ prohibits population transfer from the initially occupied Gaussian to the unoccupied ones. This leads to dynamics that are identical to the single \gls{FG} variant. 
Admittedly, the current way of initializing the wavefunction, while functional, is rather crude. Since this investigation serves more as a proof of concept than an optimized implementation, we leave further technical improvements to future work.  

\section{Conclusion}

We presented a variational hybrid \gls{CV-DV} algorithm that simulates adiabatic quantum nuclear dynamics. The algorithm embeds the nuclear wave function into the state of the oscillator-qubits register, and in effect represents it as a superposition of \glspl{FG}. The nuclear wave function is thus parameterized by qubit rotations and conditional-Gaussian gates parameters. Their time-evolution satisfy time-dependent \glspl{VP}. The number of Gaussians scales exponentially with respect to the number of ancilla qubits/hardware resources making up the circuits. 

The algorithm uses an oscillator-qubits register as a co-processor to compute matrix elements that are part of the \gls{EOM} derived from time-dependent \glspl{VP}. The matrix elements are overlaps of wave function derivatives with respect to circuit parameters. These are approximated with a finite central difference scheme by simultaneously embedding multiple Gaussian superpositions displaced along circuit parameters in the register. This approach can be viewed as a \gls{CV-DV} version of the ones presented in Refs.~\citenum{lee2022variational, ollitrault2021molecular, ollitrault2023quantum}, which use the \gls{DV} model. 

A single \gls{FG} variant of our algorithm was used to simulate the photoelectron spectrum of \ce{SO2} from physical quantum hardware. The variational parameters were computed on classical hardware, before being used as inputs on a physical quantum circuit calculating the real and imaginary parts of the autocorrelation functions. The oscillator-qubit quantum device consisted of a single-mode superconducting microwave cavity dispersively coupled to an auxiliary transmon. The resulting simulated spectrum matches well with the experimental one. Overall, it provides experimental support that certain features the proposed algorithm can be implemented on physical quantum hardware and that the variational parameters produced from the algorithm are physically reasonable.

The algorithms were tested on 1D models, namely quadratic, Morse and double-well potentials with parameters that set the energetics in the range of nuclear vibrations. The variant that evolves a superposition of \glspl{FG} converged to the numerically exact results in most cases when increasing the number of \glspl{FG}. The algorithm was able to simulate thawed Gaussian dynamics, anharmonic dynamics in a Morse potential and describe moderately well wavepacket bifurcation on a double-well potential. 

There are many possible ways to improve the algorithm, the most obvious being to resolve the limitations addressed in Sec.~\ref{sec:current-limitations}. Additionally, it might be advantageous to incorporate controlled-squeeze and non-Gaussian gates acting on the qumode. These could in principle improve the wavefunction ansatz and possibly reduce the number of ancilla qubits required for higher accuracy. Furthermore, the current algorithm needs to be extended to wavepacket dynamics in many spatial dimensions. It is where the need for a large number of Gaussians becomes imperative. The current algorithm could also be extended to nonadiabatic dynamics where Gaussians are associated with different electronic states and can transfer between them. The current algorithm can serve as a starting point for future variational \gls{CV-DV} algorithms simulating nuclear dynamics and possibly paving a way towards scalable quantum simulation of molecules. 

\section*{Code availability}

The code used to generate the data for this paper is available from the corresponding authors upon
reasonable request.

\section*{Acknowledgments}

The authors are grateful for Next Generation Manufacturing Canada (\#14234) for providing financial support. 

\begin{appendices}
\renewcommand{\thesection}{\Alph{section}.}

\section{Measuring the nuclear Hamiltonian $\mathcal{H}$}
\label{sec:H_measmt}

A conceptually simple way of obtaining $\mathcal{H}_{\text{m}}$ is by homodyne measurements, which probe the quadrature amplitudes $x$ and $p$. Given that nearly all nuclear Hamiltonians in the Born-Oppenheimer approximation can be expressed as $\hat{\mathcal{H}}=\sum_{i=1}^{N}\frac{\hat{p}_{i}^{2}}{2m_{i}}+V(\hat{\bm{x}})$, $\mathcal{H}_m$ can be directly constructed from measurements of $x$ and $p$.

Since standard homodyne measurements have low fidelity, other means of measuring $\mathcal{H}$ may be required. In some cases, it may be more feasible to retrieve $\mathcal{H}_{\text{m}}$ from Fock state measurements. This approach is particularly suitable in the scenario where the initial wavepacket is the ground state of a harmonic oscillator where the frequency is $\omega_{\text{I}}^{\text{IR}}$ and $\hat{\mathcal{H}}$ is the Hamiltonian of another harmonic oscillator where the frequency is $\omega_{\text{F}}^{\text{IR}}$. We use the superscript \text{IR} to highlight that nuclear vibrations typically fall in the infrared regime and denote the relative displacement between their minima as $r_{\text{I}}$. 

Defining $\mathcal{H}_{\text{m}}$ from Fock state measurements requires adjusting the position and the width of the qumode state at the start of the algorithm so that it reflects the initial nuclear wavepacket. In other words, the qumode state must project onto the Fock states in the same manner as the initial nuclear wavepacket projects onto the eigenbasis of $\mathcal{H}_{\text{m}}$. This can be achieved by applying a displacement gate $\hat{D}_{\text{I}}(\alpha) = \exp(\alpha\hat{a}^{\dagger} - \alpha^{*}\hat{a})$ and squeeze gate $\hat{S}_{\text{I}}(z)=\exp(\frac{1}{2}(z^{*}\hat{a}^2-z{\hat{a}^{\dagger 2}}))$, where $z$ is real, at the start of the algorithm . While either ordering is valid, the expression for $\alpha$ is simpler if the squeeze gate is applied first, producing a displaced squeezed state. 

The parameters $\alpha$ and $z$ are determined by the condition $\ket{\phi_{0}} = \hat{D}_{\text{I}}(\alpha)\hat{S}_{\text{I}}(z)\ket{0}$, where $\ket{0}$ here is the ground state of $\mathcal{H}$ and $\ket{\phi_{0}}$ is the initial nuclear wavepacket. The variance of a displaced squeezed state is $\Delta x^2 = \frac{\hbar}{2m\omega}\exp(-2z)$ and the average position is $\braket{\hat{x}} = \sqrt{\frac{2\hbar}{m\omega}}\text{Re}(\alpha)$. Setting the variance equal to that of the initial nuclear wavepacket $\Delta x^2 = \frac{\hbar}{2m\omega_{\text{I}}^{\text{IR}}}$, with $\omega = \omega_{\text{F}}^{\text{IR}}$, and solving for $z$ gives $z=\frac{1}{2}\ln(\frac{\omega_{\text{I}}^{\text{IR}}}{\omega_{\text{F}}^{\text{IR}}})$. Similarly setting $\braket{\hat{x}} = r_{\text{I}}$ and solving for $Re(\alpha)$ gives $\text{Re}(\alpha) = \sqrt{\frac{m\omega_{\text{F}}^{\text{IR}}}{2\hbar}}r_{\text{I}}$. Each measurement then yields
$\mathcal{H}_{m} = \hbar\omega^{\text{HW}}(n_{m}+\frac{1}{2})$ where $n_{\text{m}}$ is excitation number obtained on the m\textsuperscript{th} shot and $\omega^{\text{HW}}$ is the platform-dependent hardware frequency.

\section{Choosing different variational principles for norm conservation}
\label{subsec:norm-conserv}

Here, we explain why the single and multiple \glspl{FG} variants of the algorithm rely on different \glspl{VP}. We begin by reviewing the conditions that allow the McLachlan and Kramer-Saraceno \glspl{VP} to conserve the norm of the wave function~\cite{lasser2022various}.

Variational quantum dynamics constrain the wave function to a parametric form $\Psi_{\text{A}}(t)$. It belongs to a submanifold $\mathcal{M}$ of the Hilbert space within which the exact wave function evolves. For any $\Psi_{\text{A}}(t)$, there is an associated tangent space $\mathcal{T}_{\Psi_{\text{A}}(t)}\mathcal{M}$ that contains all possible ways to tangentially pass through $\Psi_{\text{A}}(t)$~\cite{lubich2005variational, lasser2022various}.

Norm conservation is a desirable attribute of approximate nuclear wavepacket dynamics. In order for the norm of $\Psi_{\text{A}}(t)$ to be conserved under variational dynamics, its parametric form  and \gls{EOM} must guarantee that 
$\frac{\partial}{\partial t}\left\langle\Psi_{\text{A}}(t)\mid\Psi_{\text{A}}(t)\right\rangle = \text{Re}\left(\left\langle\Psi_{\text{A}}(t)\mid\frac{\partial\Psi_{\text{A}}(t)}{\partial t}\right\rangle\right)= 0$. 

The \gls{EOM} derived from the McLachlan \gls{VP} satisfy $\text{Re}\left(\left\langle\delta\Psi\middle| \frac{\partial \psi_{\text{A}}(t)}{\partial t}\right\rangle\right) = \frac{1}{\hbar}\text{Im}\left(\left\langle\delta\Psi\middle|\hat{\mathcal{H}}\middle|\psi_{\text{A}}(t)\right\rangle\right)$ for any $\delta \Psi \in \mathcal{T}_{\Psi_{\text{A}}(t)}\mathcal{M}$. Because $\hat{\mathcal{H}}$ is hermitian, it follows that $\text{Im}\left(\left\langle\Psi_{\text{A}}(t)\middle|\hat{\mathcal{H}}\middle|\psi_{\text{A}}(t)\right\rangle\right) = 0$. Thus, when the parametric form is such that $\Psi_{\text{A}}(t) \in \mathcal{T}_{\Psi_{\text{A}}(t)}\mathcal{M}$, we get, 
in virtue of the McLachlan \gls{VP}, $\text{Re}\left(\left\langle\Psi_{\text{A}}(t)\middle| \frac{\partial \Psi_{\text{A}}(t)}{\partial t}\right\rangle\right)=0$, thus fulfilling the normalization condition.

The \gls{EOM} derived from the Kramer-Saraceno \gls{VP} satisfy $\text{Im}\left(\left\langle\delta\Psi\middle| \frac{\partial \psi_{\text{A}}(t)}{\partial t}\right\rangle\right) = \frac{1}{\hbar}\text{Re}\left(\left\langle\delta\Psi\middle|\hat{\mathcal{H}}\middle|\psi_{\text{A}}(t)\right\rangle\right)$ for any $\delta \Psi \in \mathcal{T}_{\Psi_{\text{A}}(t)}\mathcal{M}$. Because $\hat{\mathcal{H}}$ is hermitian, it follows that $\text{Re}\left(\left\langle i\Psi_{\text{A}}(t)\middle|\hat{\mathcal{H}}\middle|\psi_{\text{A}}(t)\right\rangle\right) = 0$. When the parametric form is such that $i\Psi_{\text{A}}(t) \in \mathcal{T}_{\Psi_{\text{A}}(t)}\mathcal{M}$, we get in virtue of the Kramer-Saraceno \gls{VP} $\text{Im}\left(\left\langle i\Psi_{\text{A}}(t)\middle|\frac{\partial\Psi_{\text{A}}(t)}{\partial t}\right\rangle\right) = 0$, which in turn implies $\text{Re}\left(\left\langle \Psi_{\text{A}}(t)\middle|\frac{\partial\Psi_{\text{A}}(t)}{\partial t}\right\rangle\right) = 0$ thereby satisfying the normalization condition.

In both variants of our algorithm, the wave function is parameterized by $\gamma_{0}$, which defines the global phase. The partial derivatives of the wave function with respect to $\gamma_{0}$ satisfies $\ket{\frac{\partial \Psi(\bm{\lambda})}{\partial \gamma_{0}}}=i\ket{\Psi(\bm{\lambda})}$. It follows that $i\ket{\Psi(\bm{\lambda})} \in \mathcal{T}_{\ket{\Psi(\bm{\lambda})}}\mathcal{M}$ with the implication that the norm is conserved when the parameters evolve according to the Kramer-Saraceno \gls{VP}. 
In the context of dynamics of a single \gls{FG}, because $\ket{\frac{\partial \Psi(\bm{\lambda_{G}})}{\partial x}}$ and $\ket{\frac{\partial \Psi(\bm{\lambda_{G}})}{\partial p}}$ are orthogonal to $\ket{\frac{\partial \Psi(\bm{\lambda_{G}})}{\partial \gamma}}$ and $\text{Im}\left(\left\langle \frac{\partial \Psi(\bm{\lambda_{G}})}{\partial \gamma} \middle| \frac{\partial \Psi(\bm{\lambda_{G}})}{\partial \gamma}\right\rangle\right)=0$, $\text{Im}\left(\bm{M}\right)$ is singular and the Kramer-Saraceno \gls{VP} cannot be used without regularization. This is why $\ket{\Psi(\bm{\lambda_{G}})}$ is propagated according to the McLachlan \gls{VP} where the norm is trivially conserved due to unitary nature of $U(\bm{\lambda_{G}})$. In the case of the superposition of Gaussians, the unitary nature of the gates are not enough to guarantee norm-conservation when following \gls{EOM} derived from the McLachlan \gls{VP} because the nuclear wave function is defined from the projection of the $\ket{0}^{\otimes N}$ state of the ancilla subsystem onto the full register state, and projections are not necessarily norm-preserving. 

\end{appendices}

\bibliographystyle{apsrev4-2}
\bibliography{hybrid_mgwp_vardyn}

\end{document}